\documentclass[twocolumn]{aastex631}
\usepackage{comment}
\usepackage[caption=false]{subfig}
\usepackage{graphicx}
\usepackage{amsmath}
\usepackage{float}
\usepackage[outdir=./]{epstopdf}

\newcommand*{\rom}[1]{\expandafter\@slowromancap\romannumeral #1@}

\newcommand{\msun}{M$_{\odot}$ }
\newcommand{\myr}{M$_\odot$~yr$^{-1}$ }

\newcommand{\lya}{Ly$\alpha$}
\newcommand{\ha}{H$\alpha$}
\newcommand{\hb}{H$\beta$}
\newcommand{\heii}{He {\sc II}}

\newcommand{\ciii}{C {\sc III}]}
\newcommand{\civ}{C {\sc IV}}
\newcommand{\nii}{[N{\sc II}]}
\newcommand{\oiii}{[O {\sc III}]}

\newcommand{\sii}{[S {\sc II}]}

\newcommand{\civlya}{\civ/\lya}
\newcommand{\heiilya}{\heii/\lya}

\newcommand{\kms}{km\,s$^{-1}$} 

\newcommand{\ferg}{erg s$^{-1}$ cm$^{-2}$ }
\newcommand{\fergarc}{erg s$^{-1}$ cm$^{-2}$ arcsec$^{-2}$}

\newcommand{\logU}{log({\sc U})}

\received{August 21, 2023}
\revised{January 7, 2024}

\submitjournal{ApJ}

\shorttitle{Environments around 3C 9 and 4C 05.84}
\shortauthors{Sabhlok et al.}

\graphicspath{{./}{./figures/}{./figures/MOS_Figs}}

\begin{document}

\title{Circumgalactic Environments around Distant Quasars 3C 9 and 4C 05.84}

\correspondingauthor{Sanchit Sabhlok}
\email{ssabhlok@ucsd.edu}

\author[0000-0002-8780-8226]{Sanchit Sabhlok}
\affiliation{Department of Physics, University of California San Diego, 
9500 Gilman Drive 
La Jolla, CA 92093 USA}

\author[0000-0003-1034-8054]{Shelley A. Wright}
\affiliation{Department of Astronomy and Astrophysics, University of California San Diego,
9500 Gilman Drive 
La Jolla, CA 92093 USA}
\affiliation{Department of Physics, University of California San Diego, 
9500 Gilman Drive 
La Jolla, CA 92093 USA}

\author[0000-0002-0710-3729]{Andrey Vayner}
\affiliation{Department of Physics and Astronomy, Johns Hopkins University, Bloomberg Center, 3400 N. Charles St., Baltimore, MD 21218, USA}

\author[0009-0003-7449-7769]{Sonata Simonaitis-Boyd}
\affiliation{Department of Physics, University of California San Diego, 
9500 Gilman Drive 
La Jolla, CA 92093 USA}

\author{Norman Murray}
\affiliation{Canadian Institute for Theoretical Astrophysics, University of Toronto, 60 St. George Street, Toronto, ON M5S 3H8, Canada}

\author[0000-0003-3498-2973]{Lee Armus}
\affiliation{IPAC, California Institute of Technology, 1200 E. California Blvd., Pasadena, CA 91125 USA}

\author[0000-0002-2248-6107]{Maren Cosens}
\affiliation{The Observatories of the Carnegie Institute for Science, 813 Santa Barbara Street, Pasadena, CA 91101}

\author[0000-0003-2687-9618]{James Wiley}
\affiliation{Department of Physics, University of California San Diego, 
9500 Gilman Drive 
La Jolla, CA 92093 USA}

\author[0000-0002-7613-9872]{Mariska Kriek}
\affiliation{Leiden Observatory, Leiden University, P.O. Box 9513, 2300 RA Leiden, The Netherlands}

\begin{abstract}
We present results from the ``Quasar hosts Unveiled by high Angular Resolution Techniques" (QUART) survey studying the Circumgalactic Medium  (CGM) by observing rest-frame UV emission lines \lya, \civ\, and \heii\, around two radio-loud quasars, 3C 9 (z=2.02) and 4C 05.84 (z=2.32), using Keck Cosmic Web Imager (KCWI). We detect large-scale Ly-alpha nebulae around both quasars with projected diameters $\sim$ 100 kpc, with spatially resolved, embedded 15-30 kpc \heii\, and \civ\, nebulae around both quasars as well as kinematically distinct \heii\, and \civ\, nebulae at a physical separation of $\sim$ 15 kpc from both quasars. Observations of \ha, \hb, and \oiii\, emission using Keck MOSFIRE spectroscopically confirm that the \lya\, nebulae extend to companion galaxies and that these quasars are in a protogroup/protocluster environment. We confirm that the \heii\, and \civ\, emission is kinematically and spatially coincident with the companion galaxies. We estimate the virial masses of the companion galaxies, their metallicities, and star formation rates, and investigate the sources of ionization. We measure the dynamical mass of the host dark matter halos and estimate that the dark matter halos of these systems will grow to a mass of $2 \times 10^{14}$ \msun (3C 9) and $2 \times 10^{13}$ \msun (4C 05.84) by z=0. The combined CGM and companion galaxies observations indicate \lya\, substructure can indicate the presence of companion galaxies in the CGM.
\end{abstract}


\section{Introduction} \label{sec:intro}

Recent theoretical and observational studies have highlighted the necessity and utility of studying the internal properties of galaxies in conjunction with their larger environment and underlying circumgalactic medium (CGM). The complex interplay between star formation \citep{Madau2014}, metallicity, dust, active galactic nucleus (AGN) and the subsequent growth of supermassive black holes (SMBH) \citep{Kormendy2013}, plays a critical role in a galaxy’s growth. Studying the connection and role of the CGM in the growth of galaxy progenitors is observationally challenging since it requires investigating the underlying physics that operates across a wide range of length scales (parsecs to 100 kiloparsecs) and time scales \citep{Tumlinson2017}. Additionally, it requires understanding the impact of potential mergers or satellite galaxies on the evolution of the central host galaxy, which is largely dominated by the gravitational interaction of dark matter halos in which galaxies reside \citep{Overzier2016}.\\ 
Simulations have fundamental issues trying to replicate physics across a dynamic range of length scales due to limited numerical resolution. In practical terms, this means implementation of subgrid models for numerically unresolved but important phenomena such as Molecular Clouds (O(10-100 pc)), Supernovae or AGN. On larger length scales, cosmological simulations tend to focus on Dark Matter (DM) only simulations making it difficult to study the impact of baryonic physics on large scale structure. Recently, multiple simulations have studied a numerically resolved model of the CGM, investigating the implications of enhanced spatial resolution on the convergence of observable properties of the CGM \citep{vandeVoort2019, Peeples2019FOGGIE, Hummels2019, Bennett2020, Gronke2022}. They find that enhanced spatial resolution increases total H {\sc I} covering fraction, increased cold gas content and increased inflowing gas mas flux. These simulations show that cloud sizes are still not converged and improving the spatial resolution of these simulations could allow for larger observed values of cold gas mass \citep{Cantalupo2014}. Additionally, spatial resolution of the CGM is important to match the observed distribution of absorber properties \cite{OMeara2015} and that spatial resolution can affect the predicted total luminosity of \ha\, by a factor of two and by an order of magnitude for other lines such as C {\sc III} and C {\sc IV} \citep{Corlies2020}, something only observations can clarify. Thus, it is critical to complement our understanding of galaxy evolution gained from simulations with multi-wavelength observations of individual sources to connect the dots between physical processes occurring across a large dynamical range of length scales. \\
On Keck, the Multi-Object Spectrometer for Infra-Red Exploration (MOSFIRE) has enabled targeted surveys such as the  MOSFIRE Deep Evolution Field Survey (MOSDEF) \citep{Kriek2015} and the Keck Baryonic Sky Survey (KBSS) \citep{Steidel2014KBSS} which build upon the success of the Sloan Digital Sky Survey (SDSS) \citep{York2000SDSS} by extending our understanding of stellar evolution, metallicities and star formation to higher redshifts. Instruments such as the Keck Cosmic Web Imager (KCWI) \citep{Morrissey2018KCWI} on Keck and the Multi Unit Spectroscopic Explorer (MUSE) \citep{Bacon2010MUSE} on the Very Large Telescope (VLT) have come online to present compelling pictures of the CGM surrounding galaxies at $1.5 < z < 4$. Prior to the advent of these instruments, narrow band imaging revealed some of the first observed \lya\, blobs \citep{Hu1987,Steidel2000, Heckman1991a}, with follow up long slit spectroscopy revealing spatially extended emission 10s of kpc away \citep{Heckman1991b}. Theoretical work indicated extended \lya\, emission was linked to the multiphase structure of infalling CGM gas \citep{Haiman2001}. Additionally, these nebulae were observed in the vicinity of quasars since surface brightness of the nebulae is boosted by the fluorescence emission due to recombination radiation \citep{Cantalupo2005,Kollmeier2010}. The advent of the Integral Field Spectrographs (IFS) KCWI and MUSE means that instead of narrow band imaging followed by long slit spectroscopy, we can now spatially map out the emission and kinematics of such nebulae around potential targets with greater ease. \cite{Borisova2016} identified 17 radio-quiet quasars in addition to 2 radio loud quasars with extended \lya\, emission using MUSE/VLT finding cold gas ubiqitous around these sources within 50 kpc at $3 < z < 4$. \cite{AB2019} identified a sample of 61 \lya\, nebulae around quasars, 15 of which are radio loud, with MUSE/VLT at $z\sim3$, studying the morphology and sources of ionization for the nebulae. \cite{Cai2019} studied a sample of 16 ultraluminous Type 1 QSOs around $z \sim 2$ using KCWI and suggested that cool gas mass decreases going from $z \sim 3$ to $z \sim 2$. Recent works have investigated the role of CGM around radio-loud AGN using \lya\, spectroscopy finding evidence for correlation between \lya\, detection and the extent of the radio source \citep{Shukla2022} and further evidence of companion galaxies around these sources \citep{Vayner2023}. While these observations have established the ubiquity of \lya\, nebulae and their prevalence around quasars, separating the role of the host galaxies in the evolution of these nebulae from their surroundings requires us to study them in tandem with a survey of the larger environment.\\ 
Tracing the origins of the most massive clusters remains challenging at high-redshift since identifying gravitationally bound structures requires spectroscopic confirmation. The Visible Multi-Object Spectrograph (VIMOS) Ultra Deep Field (UDF) Survey \citep{LeFevre2005} surveyed 0.61 squared degrees of the night sky using the VLT and identified several large protoclusters of galaxies \citep{Cucciati2014, Lemaux2014, Cucciati2018}. However, in the absence of a large spectroscopic survey, searches for these structures need to be conducted by spectroscopic follow-up on imaging data sets. \cite{Wylezalek2013} used \textit{Spitzer} imaging to search for galaxy clusters around radio-loud AGN at $1.3 < z < 3.2$. This study observed an overdensity of sources around these AGN with Spitzer Infrared Array Camera (IRAC) imaging, suggesting that radio-loud quasars can be used to look for high redshift protoclusters. These protoclusters are likely to be progenitors of the Brightest Cluster Galaxies (BCG) found at the centers of the largest clusters observed in present-day. \\
Since the CGM is a large, diffuse, multiphase medium, having as much information about the system being studied would help decipher the impact of different processes on the CGM. Selection of radio loud quasars to study the CGM around the host galaxies offers several advantages. Firstly, the quasar illuminates the CGM and boosts \lya\, fluorescence which makes observing large \lya\, nebulae around these sources easier. Secondly, in an idealized version of the unified AGN model, the orientation of the radio jet is expected to be normal to the accretion disk and the surrounding dusty torus in an AGN. Since we primarily expect \lya\, to be emitting due to quasar photoionization and due to scattering of light from the AGN and/or the host galaxy, the jet provides a valuable reference orientation against which \lya\, morphology can be compared. Lastly, literature indicates that radio-loud AGN at high redshift are some of the rarest objects in the universe and that they preferentially reside in protocluster environments \citep{Rigby2011, Wylezalek2013, Hatch2014, Lacy2020}. Observing the CGM around radio-loud quasars could potentially reveal signatures of companion galaxies, residing in the quasar host galaxy's dark matter halo. Previous detection of large \civ\, and \heii\, emission around quasars have provided valuable information regarding the ionizing source of the gas, which has been attributed to star formation \citep{VillarMartin2007a} and quasar ionization \citep{Falkendal2021}. \civ\, and \heii\ lines are smaller in extent than \lya\, emission and found to be embedded within the CGM halos of the host galaxy \citep{VillarMartin2003, Vernet2017}. Thus, mapping the extent and relative strengths of these lines with respect to \lya\, may provide valuable insight regarding the ionizing source powering the emission. \\
In this paper, we present new results of two well studied radio-loud quasars, 3C 9 ($z=2.02$) and 4C 05.84 ($z=2.32$) using new observations in conjunction with a wealth of ancillary data, described in detail in Section \ref{sec:TarSel}. Our targets provide a unique opportunity to study the interplay of galactic outflows, radio jets, CGM, and companion galaxies. \\
This paper is organized as follows. In section \ref{sec:TarSel}, we describe the target selection, ancillary data and observations. In section \ref{sec:ODR}, we discuss the data reduction techniques employed. In section \ref{sec:CGM}, we look at the PSF subtracted \lya, \heii\, and \civ\, emission around the quasars. In section  \ref{sec:MOSFIRE} we discuss the confirmed spectroscopic protogroup/protocluster members in the larger environments around these quasars. In section \ref{sec:Discussion}, we discuss our findings, considering the impact of the environment on the CGM, sources of ionization powering the nebulae, the mass growth of the two quasar systems, and the inner and outer CGM model as applicable to our observations. We summarise our findings in section \ref{sec:Summary}. Unless stated otherwise, we assume $\Omega_M = 0.308$, $\Omega_{\Lambda} = 0.692$ and $H_0 = 67.8 \, \mathrm{km \, s}^{-1}\, \mathrm{Mpc}^{-1}$ for the cosmological parameters \citep{Planck2016}.

\section{Target Selection} \label{sec:TarSel}
Herein, we present new results on two sources, radio-loud quasars 3C 9 (z=2.02) and 4C 05.84 (z=2.32). These quasars as part of the "Quasar hosts Unveiled by high Angular Resolution Techniques" (QUART) survey \citep{QUART3C298,QUART1,QUART2} initially selected for observations with Keck "OH-Suppressing Infra-Red Imaging Spectrograph" (OSIRIS) \citep{OSIRIS2006} and the facility laser-guide star adaptive optics system (LGS-AO). 3C 9 is one of the most luminous sources in the 3C radio catalog. \cite{Bridle1994} first observed the detailed radio jet and lobe structure around the quasar using deep VLA observations, whereas \cite{Wilman2000} first detected extended emission lines around the quasar. Chandra observations of 3C 9 revealed extended X-ray emission from the quasar with two sidelobes aligned with the radio lobes observed using VLA. 4C 05.84 was first identified by \cite{Gower1967} as a radio source. \cite{Barthel1988} carried out detailed radio observations establishing the extent and geometry of the radio jet and the spectral index for the quasar. \cite{Barthel1990} measured the redshift of the source as $2.323$ based on an average of observed emission lines {($\mathrm{\lambda_{Ly\alpha}} = 1215.67 \mathrm{\AA}; \, \mathrm{\lambda_{C IV}} = 1549.06 \mathrm{\AA}; \, \mathrm{\lambda_{C III}} = 1908.73 \mathrm{\AA}$)}. \cite{Heckman1991a} carried out deep optical imaging of both quasars and detected extended \lya\, nebulae around both 4C 05.84 and 3C 9. \\
OSIRIS observations of the quasars' host galaxies revealed that either star formation or the AGN could drive outflows in 3C 9, since they are aligned with the radio jet of the quasar but appear to originate in the star forming regions without extending to the quasar host galaxy nucleus. For 4C 05.84, observations revealed the presence of galactic scale outflows most likely driven by either an isothermal shock or due to radiation pressure by the quasar. Follow up ALMA observations of 4C 05.84 \citep{QUART_ALMA} reveal that the outflow in the quasar host galaxy is multi gas phase with the detection of a molecular outflow traced using the CO (4-3) transition. While the majority of the energetics are in the ionized gas phase, most of the mass in the outflow is found in the cold molecular gas phase. The detected outflows are also aligned with the path of the jet indicating that the jet could be responsible for driving the outflow. We found that both quasar host galaxies lie off the local scaling relationship between the mass of the Supermassive Black Hole (SMBH) and the mass of the galaxy \citep{McConnell2013}. The stellar mass of both galaxies will require growth by a factor of $ \sim 10$ between redshift of 2 and 0 in order to end up on the local scaling relationship.\\
To target potential companion galaxies around both quasars using MOSFIRE, we used Spitzer and Hubble imaging to identify potential targets in the environment around the quasars for spectroscopic confirmation and optical nebular line studies. We used publicly available observations from the Hubble Space Telescope (HST) WFC3 and ACS instruments, along with data from the Spitzer IRAC 3.6 $\mu$m and 4.5 $\mu$m channels to perform an overdensity analysis. All the {\it HST} data used in this paper can be found in MAST: \dataset[10.17909/czbk-dw64]{http://dx.doi.org/10.17909/czbk-dw64}. The Spitzer data products were obtained from \cite{https://doi.org/10.26131/irsa3}. {We have utilized HST WFC3 IR F160 images for 3C 9. No HST data was available for 4C 05.84 at the time of analysis.}

\begin{figure*}
    \centering
    \includegraphics[width=\linewidth]{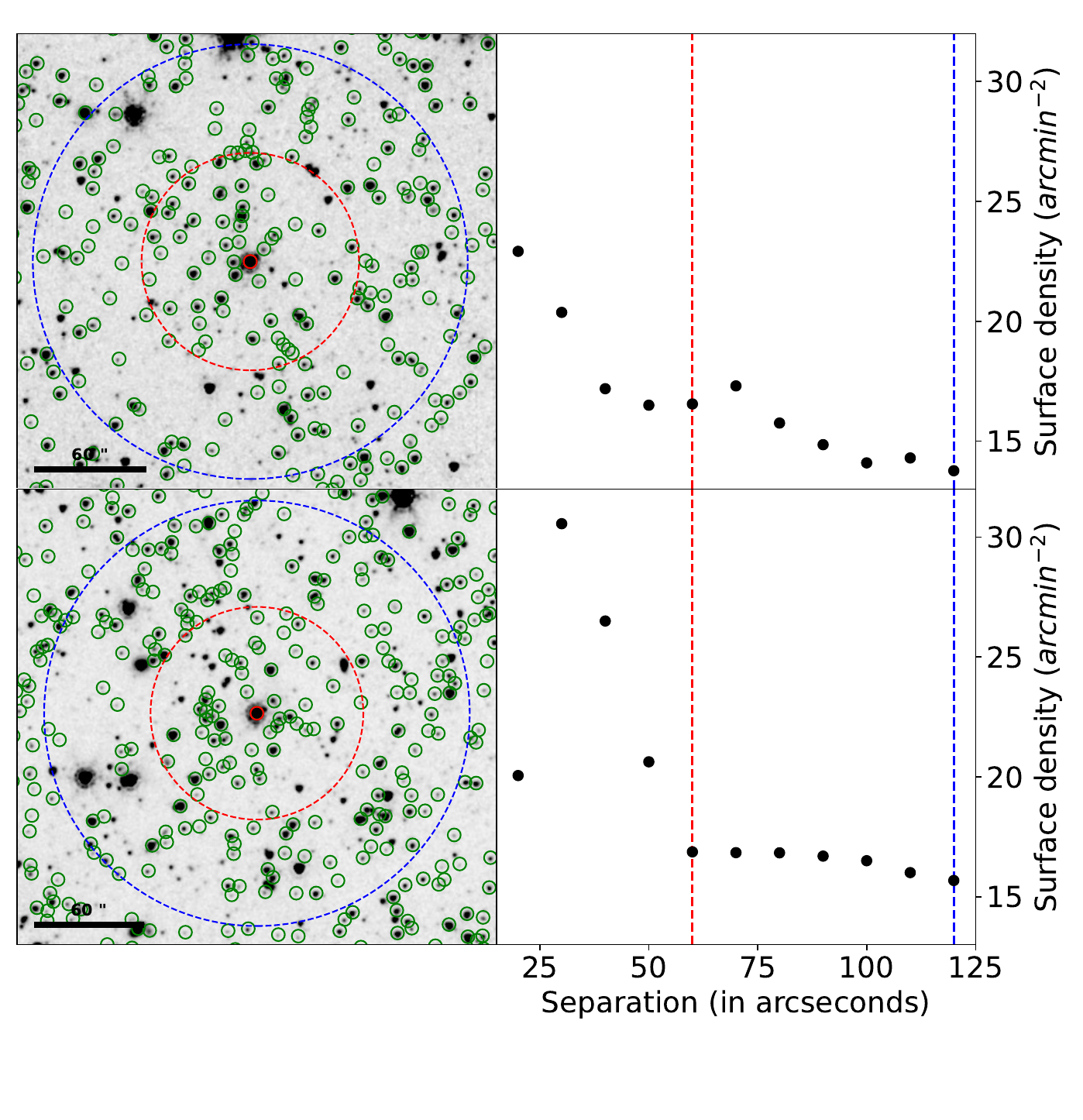}
    \caption{(LEFT) Overdensities measured around the quasars 3C 9 (Top) and 4C 05.84 (Bottom) from archival Spitzer data. The 4.5 $\mu m$ IRAC image centered on the quasar (circled in red) is shown in both cases. We perform a color cut $[3.6]_{AB} - [4.5]_{AB} > - 0.1$ \citep{2008ApJ...676..206P} to select sources with $z>1.3$ (circled in green). We plot $1 \, arcmin$ and $2 \, arcmin$ circles centered on the quasar in dashed red and blue lines respectively. (RIGHT)The surface density of sources (in sources $arcmin^{-2}$) against separation from the quasar (in $arcseconds$). The vertical dashed red line at $60 \arcsec$ is shown to indicate the number density of sources used for comparison against the SpUDS blank field number density $\Sigma_{\text{SpUDS}} = 8.3 \, \text{arcmin}^{-2}$ in \citet{Wylezalek2013}. Both quasars show a factor of $\sim$ 2 increase in the number of sources at this radius compared to the field.} 
    \label{fig:spitzer_colorcut}
\end{figure*}

We followed the procedure outlined in \citet{Wylezalek2013} to analyse the Spitzer archival data. We performed a color cut $[3.6]_{AB} - [4.5]_{AB} > - 0.1$ to select sources with $z > 1.3$ \citep{2008ApJ...676..206P}. We present the results of the color cut in Figure \ref{fig:spitzer_colorcut}. We calculate the number density of such sources around 3C9 in a 1 arcmin radius to be $\Sigma_{3C9} = 16.5 \, \text{arcmin}^{-2}$ which compared to the Spitzer UKIDSS Ultra Deep Survey (SpUDS) blank field {($\Sigma_{\text{SpUDS}} = 8.3 \, \text{arcmin}^{-2}$; $\sigma_{\text{SpUDS}} = 1.6 \, \text{arcmin}^{-2}$)} in \citet{Wylezalek2013} implies an overdensity of $5.1\sigma$. Similarly, we obtain $\Sigma_{4C0584} = 16.8 \, \text{arcmin}^{-2}$ which implies an overdensity of $5.3\sigma$. The exposure times for archival data were shorter than those used by \citet{Wylezalek2013} and therefore result in relatively higher errors for IRAC fluxes obtained from Spitzer. We note that these errors propagate into our color calculations, which could increase the number of outliers. 

\section{Observations \& Data Reduction} \label{sec:ODR}
\subsection{NIRC2}
We have archival NIRC2 observations of the quasar 4C 05.84 (PI: Stockton), which we have reduced using in house code to obtain the RGB composite shown in Figure \ref{fig:MOS_Slits}. Our observations consist of AO Laser Guide Star (LGS) assisted imaging of the quasar in the J, H and Kp bands on NIRC2{, using the wide camera at a plate scale of 0.04 $\arcsec$/pixel}. The total exposure times were 25 minutes in Kp, 25 minutes in H and 54 minutes in the J band. Our data reduction follows the procedure outlined in \cite{Melbourne2005} using the distortion solution from \cite{Yelda2010}. {We calculate sky residual brightness to be 22.2 mag arcsec$^{-2}$ for J band, 21.2 mag arcsec$^{-2}$ for H band and 20.3 mag arcsec$^{-2}$ for Kp band.} We have combined the data into an RGB using the Astropy implementation of the method outlined by \cite{Lupton2004}. \\

\subsection{KCWI}
We observed 3C9 using KCWI on October 3/4, 2018 and followed up on August 17, 2020. The total time spent on 3C 9 was 6.5 hours. We observed 4C 05.84 with KCWI on November 22/23, 2017, July 8-10 2018 and October 2, 2018, for a total of 9.3 hours. {We dithered with an ABAAB pattern where the A frame targets the quasars and the B frames are ``pure" sky frames, ensuring we have at least one neighboring sky frame for each object frame. For both quasars, we used the medium slicer with the BL grating and KBlue filter with the wavelength centered at 4500 \AA. For the final reduced data, our wavelength coverage was $3532\AA$ to $5527 \AA$, at a spectral resolution of 1,800. The field of view for this KCWI configuration is $16.5 \textrm{\arcsec}\, \times 20.4 \textrm{\arcsec}$, which was spatially dithered in units of $0.3 \arcsec$.} We observed standard stars at the beginning and end of the night. 

Primary data reduction was carried out using the KCWI Data Reduction Pipeline v1.1 \citep{KCWIPipeline}. The pipeline performs cosmic ray removal, wavelength solution, flat-fielding and dark corrections, differential atmospheric refraction corrections and flux calibration. We skipped the pipeline sky subtraction and performed sky subtraction using our own routine. Sky subtraction was carried out by comparing the observed average flux in a sky patch on the object frame with a pure sky frame to obtain a scaling factor. Since this scaling factor may vary with wavelength, we performed this piece-wise along the wavelength axis to get the best residuals post sky subtraction. To correct the WCS information in the sky subtracted frames, we performed a 2D Gaussian fit using the photutils package to determine the quasar centroid. This Gaussian fit was performed at a wavelength slice far from the quasar expected emission lines to sample the point spread function (PSF). We assigned this centroid the coordinates from GAIA \citep{GAIARef}. These WCS corrected, sky subtracted cubes were co-added using CWITools \citep{CWIToolsRef} to give the final reduced datacube. The package also re-samples the datacubes from their native pixel scale of $0.69\arcsec \times 0.29\arcsec$ to a grid with a pixel scale of $0.3\arcsec \times 0.3\arcsec$ and combines individual data cubes using the Drizzle algorithm \citep{Fruchter2002}. After performing PSF Subtraction on this co-added datacube, we then used CWITools to obtain segmentation maps for \lya, \heii\, and \civ\, emissions lines, as well as subsequent surface brightness and moment maps for each emission line. \\
We performed PSF subtraction on the KCWI data as described in \citet{QUART_OSIRIS}. To summarise the procedure, we construct a PSF image by using the continuum emission from the quasar and the wings of the broadline emission lines at large velocity offset from systemic to avoid including any potential extended emission from the host galaxy or the CGM. This PSF image is then normalized to the peak flux and subtracted out at each wavelength channel while re-scaling the flux to the peak of the emission at that channel. To verify good spatial and spectral PSF subtraction, we proceed as follows. We analyse the radial profile of the resultant PSF subtracted data cube and compare it directly to the original data. For good PSF subtraction, we expect the flux to become comparable to sky residuals at a distance comparable to the size of the seeing disk, which for our nights was 0.5 - 0.7\arcsec. The reconstructed PSF FWHM from KCWI data (with a slight wavelength dependence) is well-matched. There are wavelength dependencies to the PSF and spaxels close to the quasar show residual broad-line emission. PSF subtraction is good for all spaxels further than 0.6\arcsec - 0.9\arcsec (2 - 3 pixels) of the quasar. To optimize PSF subtraction we have constrained our PSF to two critical wavelength regimes, those around the \lya\, and \civ\, emission lines, constraining the broad line residual emission and over subtraction to within the expected sky residual noise. Surface brightness maps beyond 0.9\arcsec (3 pixels) should be minimally effected by PSF subtraction. \\

\subsection{MOSFIRE}
We observed 3C 9 using MOSFIRE on August 11/12, 2020 and 4C 05.84 on August 13, 2020 in H and K bands. The weather was clear with seeing estimated between 0.4$\arcsec$ and 0.5$\arcsec$. We aimed to detect \ha,\nii, \sii\ in the K-band and \oiii\, and \hb\, in the H-band, given that 3C 9 is at a redshift of 2.019 and 4C 05.84 at 2.32. The slitmask configurations are described in Table \ref{tab:Slitmasks}. For each slitmask, we observed one continuum source with H magnitude less than 15 in a science slit to correct for instrumental drift during reductions. Other than Mask 1 which used the 3C 9 quasar as the continuum source itself, all other masks used stars from The Two Micron All Sky Survey (2MASS) \citep{2MASS} as continuum sources. We used a slit width of 0.7$\arcsec$ which gives $R = 3270$. The individual exposure time in H band was 120s whereas for the K band it was 180s. We dithered using the ABBA observing mode. We observed the star HIP 116886 (spectral type A0V), selected from the ESO list for Spectrophotometric standards,
\footnote{www.eso.org/sci/observing/tools/standards/spectra/stanlis} for telluric correction and flux calibration using the long2pos slitmask.

\begin{deluxetable}{cccccc}

\caption{MOSFIRE Slitmask configurations \label{tab:Slitmasks}}
\tablenum{1}
\tablehead{\colhead{Quasar} & \colhead{Mask} & \colhead{PA} & \colhead{Targets} & \colhead{Exp Time} & \colhead{Band} \\ 
\colhead{} & \colhead{} & \colhead{degrees} & \colhead{\#} & \colhead{min} & \colhead{} } 
\startdata
3C 9 & 1 & -8.0 & 35 & 72 & H \\
3C 9 & 1 & -8.0 & 35 & 56 & K \\
3C 9 & 2 & 53.0 & 32 & 36 & H \\
3C 9 & 3 & 108.0 & 30 & 72 & H \\
4C 05.84 & 1 & 7.0 & 30 & 38 & H \\
4C 05.84 & 2 & 102.0 & 30 & 38 & H \\
4C 05.84 & 2 & 102.0 & 30 & 36 & K \\
\enddata
\end{deluxetable}

We reduced the data using the MOSDEF pipeline \citep{2015ApJS..218...15K}. The pipeline carries out sky subtraction, masks bad pixels and cosmic rays, and combines individual frames. The pipeline also corrects for instrumental drift using the specified continuum source in a mask. We used the observed telluric star to perform our own flux calibration. We first perform a wavelength calibration for the telluric spectrum using arc lamps in K-band and atmospheric OH lines in H-band. We extracted the 1D spectrum of the star observed using the long2pos\_specphot mask. The long2pos script obtains 4 spectra at the spectral resolution of the science exposure, two at the low end and two at the high end of wavelength coverage. We pick the higher throughput spectrum from each wavelength end. We fit a Gaussian to any hydrogen absorption lines in the stellar spectrum to subtract them out. We also detect and remove spurious spikes in the extracted spectrum that are from residual bad pixels. We then construct a blackbody emission model for the effective temperature of the spectral type, normalised to the 2MASS flux of the observed standard\citep{2MASS}. We divide the stellar spectrum by the blackbody spectrum to get the response curve for the instrument, which includes atmospheric and telescope-instrument response. We use the response spectrum to convert the extracted spectra from $Datanumbers/s$ to $ergs \, cm^{-2} \, s^{-1}$/\AA. Lastly, we independently checked the quality of flux calibration by checking the flux of continuum sources on science masks and comparing them to the expected 2MASS flux densities. {We find that the MOSFIRE derived magnitudes are in agreement to within 0.2 mag for all H and K band masks, after correcting for slit loss and the differences between MOSFIRE and 2MASS bandpasses. Since \hb\, emission for 3C 9 falls close to the atmospheric window, the flux calibration for \hb\, has higher uncertainty.} We also compare the response spectrum across different nights and find them to be qualitatively consistent with each other, {i.e. the wavelengths for observed sky lines and any absorption lines in the telluric stars are consistent across different nights}.\\

\begin{figure*}
    \centering
    
    \includegraphics[width=\linewidth]{ 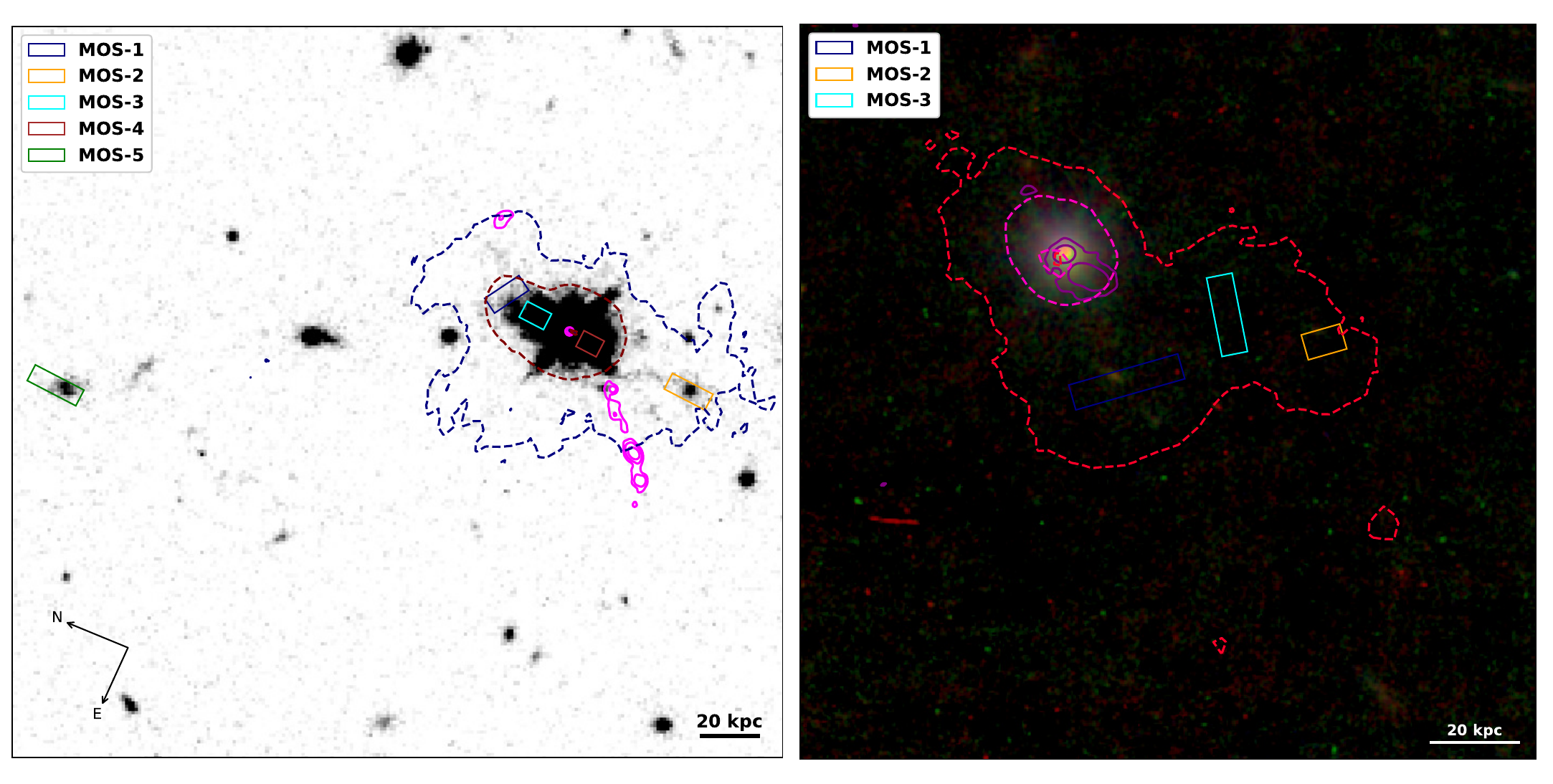}
    
    \caption{We show the \lya\, nebulae as seen by KCWI, shown as dashed contours for both quasars. The inner and outer contours correspond to an observed flux density of 1.0 and 0.1 $\times 10^{-16}${\fergarc} respectively. For 3C 9, we overlay this on an HST WFC3 F160 image (Left) and for 4C 05.84 we overlay MOSFIRE slits on a NIRC2 RGB composite where the red, green and blue channels correspond to Kp, H and J band images of the quasar 4C 05.84 respectively (Right). We also display the spatial location of emission lines as detected with MOSFIRE and label the individual sources as shown in Table \ref{tab:Group-members-coords}. We also overlay the contours of the radio jets coming out of the two quasars in {magenta}.}
    \label{fig:MOS_Slits}
\end{figure*}


\section{KCWI Results: Circumgalactic Medium} \label{sec:CGM}

\begin{figure*}[!ht]
    \centering
    \includegraphics[width=\linewidth]{ 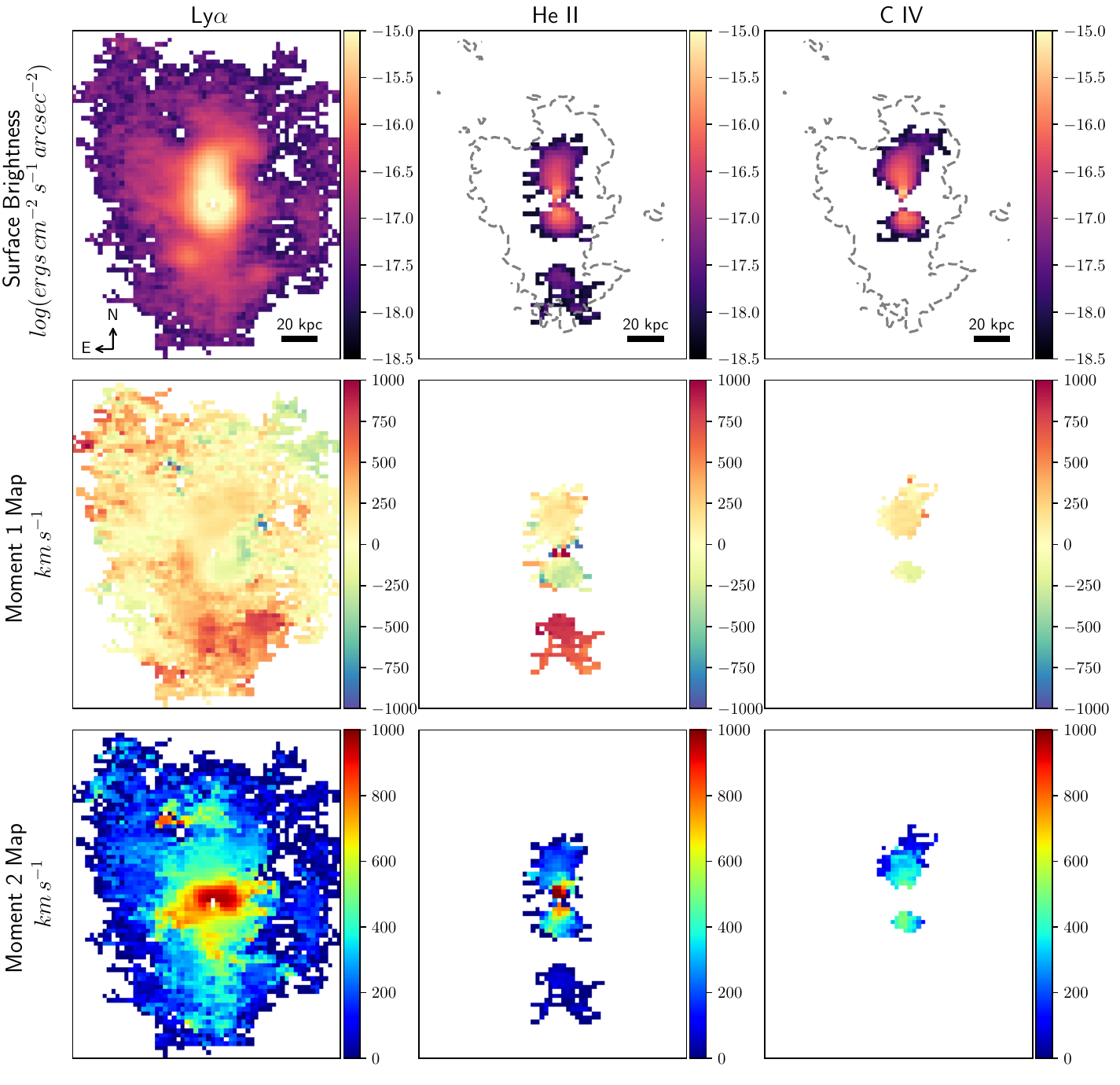}
    \caption{A PSF-subtracted image of the \lya\,(Left column), \heii\,(Middle column) and \civ\, (Right Column) derived properties for 3C 9. The top row is the surface brightness, middle row is the moment 1 map, and bottom row is the moment 2 map. All figures have an on-sky position angle of 0. The systemic redshift used to calculate the moment maps is used from OSIRIS measurements in \citet{QUART_OSIRIS}. {The grey dashed contour in \heii\, and \civ\, Surface Brightness figures denotes a \lya\, isophote at a surface brightness of $10^{-17}$ \fergarc.} }
    \label{fig:KCWI-3C9}
\end{figure*}

\begin{figure*}[!ht]
    \centering
    \includegraphics[width=\linewidth]{ 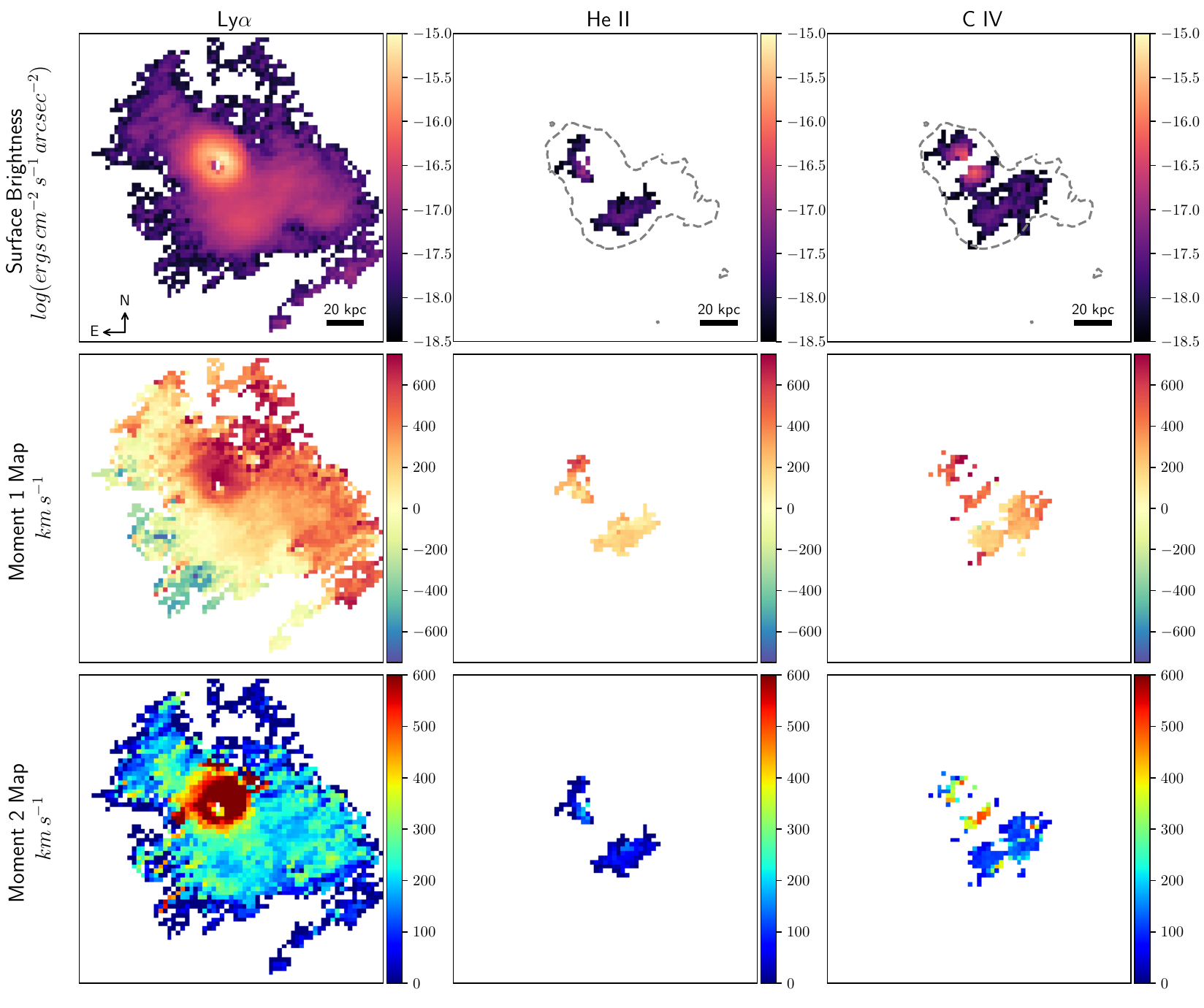}
    \caption{A PSF-subtracted image of the \lya\,(Left column), \heii\,(Middle column) and \civ\, (Right Column) derived properties for 4C 05.84. Same units and PA configuration as Figure \ref{fig:KCWI-3C9}.}
    \label{fig:KCWI-4C0584}
\end{figure*}

We extract flux and moment maps of each emission line for 3C 9 and 4C 05.84 using the CWITools package. We first perform a segmentation of the PSF-subtracted data cube for each emission line (\lya, \civ\, and \heii), using a minimum threshold of $2\sigma$, using the $cwisegment$ routine. This gives us a resultant masked datacube where all pixels below the minimum threshold are masked. A zero-moment map is created by simply summing the flux from all the unmasked spaxels. The velocity (moment 1) and dispersion (moment 2) maps are created similarly by applying the respective moment equations to the unmasked spaxels in this datacube. The resultant moment maps for 3C 9 are in Figure \ref{fig:KCWI-3C9} and 4C 05.84 are in Figure \ref{fig:KCWI-4C0584}.  

\subsection{CGM Morphology \& Kinematics}

\begin{figure*}[!ht]
    \centering
    {{\includegraphics[width=\linewidth]{ 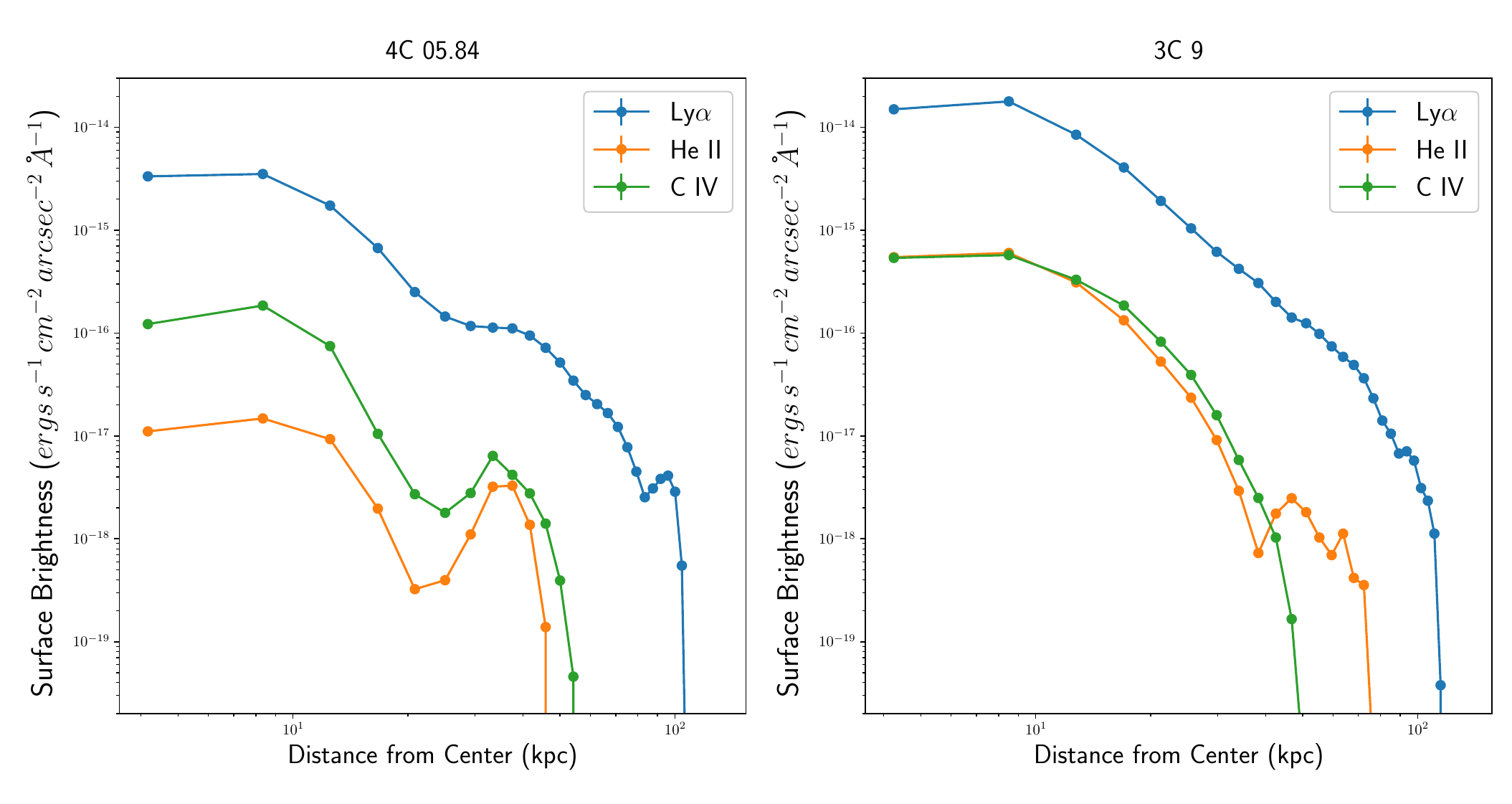} }}%
    \caption{Radial surface brightness profiles of \lya, \civ\, and \heii\,derived from 2d surface brightness maps: see Figures \ref{fig:KCWI-3C9} and \ref{fig:KCWI-4C0584} for 4C 05.84 (Left) and 3C 9 (Right). We center the surfaces on the quasar and construct successively larger annular regions to extract the azimuthally averaged surface brightness. We omitted the innermost 2 pixels, which corresponds to 0.6 \arcsec corresponding or approximately 5 comoving kpc to avoid any contamination from PSF subtracted residuals.}%
    \label{fig:Q_Radial_Profile}
\end{figure*}

\subsubsection{3C 9 CGM}

The KCWI observations are sensitive to 2$\sigma$ flux limits of {$5 \times 10^{-20}$ \fergarc}. We detect \lya\,emission extending throughout our field of view around 3C 9 with the most prominent emission seen as a nebulae with a diameter of $~ \, 120 \, kpc$ centered on the quasar. The southern part of the nebula shows distinct clumps of \lya\, with two faint peaks clearly identified southwest and southeast of the quasar. We also detect two distinct \heii\, nebulae, the first around the quasar host galaxy with a spatial extent of $~ \, 40-50 kpc$ and a second about $30 \, kpc$ away, with a spatial extent of $\sim$ $30-40 kpc$. Both \heii\, and \civ\, emission around the quasar appears to have a biconical morphology. The radial profile of the three emission lines for the two quasars are in Figure \ref{fig:Q_Radial_Profile}. For purposes of discussion, we can demarcate different regions in the CGM around the quasars as shown in Figure \ref{fig:Q_Ratios_im}. We have region \textit{3C9-a} (Northern cone), region \textit{3C9-b} (Southern cone) and region \textit{3C9-c} (Southern cloud further away from the quasar). We detect a \civ\, emission nebula coincident with the \heii\, emission closer to the quasar in \textit{3C9-a} and \textit{3C9-b}, but do not detect any spatially resolved \civ\, emission coincident with the \heii\, nebula in \textit{3C9-c}. We note that \heii\, emission does not necessarily coincide with peaks in \lya\, surface brightness, in fact, we detect no \heii\, for the bright {south-east} \lya\, peak or the fainter {south-west} \lya\, peak. 

As discussed in section \ref{sec:ODR}, we have residual emission from the broad line region in the innermost 2-3 pixels, which translates to {$\sim 7$ kpc} from the center. As a result, we treat the innermost radial profile as an upper limit and discuss the features seen greater than {7 kpc} away, which corresponds to 3 pixels. The radial profiles for 3C 9 similarly indicate that the \heii\,emission seen in region \textit{3C9-c} is a distinct component, showing no spatially resolved \civ\, emission, however this may be due to observational sensitivity.

        We detect a smooth velocity gradient across the nebula with velocity ranging from $- 800$ \kms\, to $1000$ \kms\, relative to the systemic redshift of the system ($z_{3C9} = 2.02$). region \textit{3C9-c} has a velocity greater than $800$ \kms\, with respect to the quasar, seen both in \heii\, and \lya. We can also see a region of high dispersion south of the quasar, with dispersion values measured to be in excess of {$500$ \kms} in \lya\, as seen in the moment 2 map in Figure \ref{fig:KCWI-3C9}. The \heii\, nebula around the quasar host galaxy shows a smooth velocity gradient from \textit{3C9-b} to \textit{3C9-a}, with a velocity gradient between $- 400$ \kms\,  to $250$ \kms. On the other hand, region \textit{3C9-c} shows a relative velocity gradient from $\sim \, 400$ \kms\, to $800$ \kms, with the brightest part of the emission in a narrow $\sim$ 15 kpc clump of gas at a velocity of $800$ \kms\, with respect to the quasar. The southern \heii\,emission has very low velocity dispersion across the nebula with values $< 100$ \kms. The emission of \heii\, in \textit{3C9-c} seems kinematically consistent with the \lya\, emission in the region, even though the \lya\, emission shows shows a gradient around the region. 

\subsubsection{4C 05.84 CGM}

We detect \lya\, emission for 4C 05.84 extending throughout our field of view up to a scale of 120 kpc in diameter in the immediate vicinity of the quasar and in faint but spatially resolved peaks {southwest} of the quasar. The presence of substructure can further be seen in the radial profile of the \lya emission in Figure \ref{fig:Q_Radial_Profile}. We detect two distinct \heii\, nebulae, the first immediately around the quasar, with an extent of 20 kpc and a second about 20 kpc {southeast} of the quasar, with an extent of 30 kpc. Similar to 3C 9, we denote different regions around the CGM as region \textit{4C0584-a} (North {Eastern} cone), region \textit{4C0584-b} (South {Western} cone) and the region \textit{4C0584-c} (South {West} cloud further away from the quasar) as seen in Figure \ref{fig:Q_Ratios_im}. The \heii\, emission in the immediate vicinity of the quasar in regions \textit{4C0584-a} and \textit{4C0584-b} is markedly fainter than that in \textit{4C0584-c}, which is also larger in size. This is surprising since we a priori expect both nebulae to be photoionized by the quasar. We detect slightly larger \civ\, emission nebulae around both \heii\, nebulae.  The \civ\, emission around the \heii\, in \textit{4C0584-c} also appears to have a biconical morphology. From the radial profile in Figure \ref{fig:Q_Radial_Profile}, we see that \heii\, emission follows that of \civ\, for 4C 05.84, including that in \textit{4C0584-c}.  

The \lya\, moment 1 map for 4C 05.84 shows a smooth gradient across the nebula ranging from $-600$ \kms\, to $800$ \kms\, with dispersion $< 400$ \kms\, for the bulk of the nebula beyond the immediate vicinity of the quasar. The \heii\, emission in \textit{4C0584-c} lies within a velocity gradient of $100$ \kms\, to $300$ \kms\, with dispersion $< 150$ \kms. Similar to 3C 9, the \heii\, emission velocity gradient seems to match the corresponding \lya\, velocity gradient in the region indicating the emission originates from the same region of gas with similar sources of ionization. 

The moment 1 map for \textit{4C0584-c} {indicates a gradual velocity change in the observed \civ\, emission from {southeast to northwest}}. However, we do not see clear evidence of a radially decreasing dispersion profile, as seen around the quasar in \civ. If the apparent biconical nature of \civ\, is due to the presence of an obscured AGN, we would expect to see higher values for velocity dispersion in the center of the nebula as a result of outflows from the AGN. Since we do not see any evidence of outflows, the apparent biconical nature of the observed \civ\, emission is likely due to the complex geometry of the nebulae and the presence of dust between the quasar and the nebula illuminating certain parts of the nebula more than others. 

\subsection{\civlya\, and \heiilya\, ratio maps} \label{subsec_CIV_HeII}
\begin{figure*}[!ht]
    \centering
    {{\includegraphics[width=\linewidth]{ 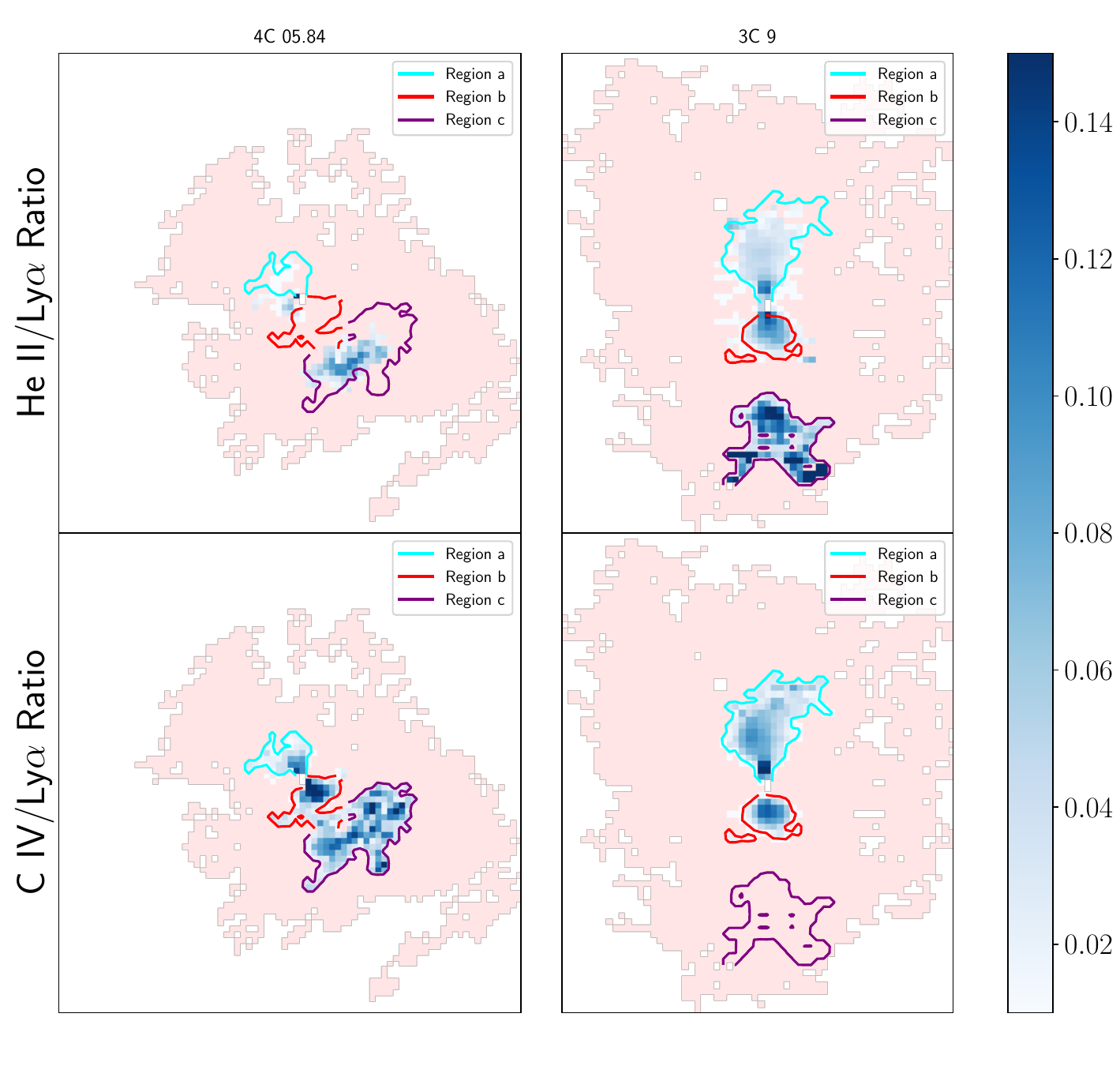} }}%
    \caption{Emission line ratios constructed from Figures \ref{fig:KCWI-3C9} and \ref{fig:KCWI-4C0584}. We divide \civ\, and \heii\, surface brightness maps by the \lya\,maps to obtain the surface brightness ratios presented here. Three distinct regions are identified based on their spatial locations. 3C 9 shows a biconical outflow in \civ\, and \heii\, with the northern cone denoted as region \textit{3C9-a} (outlined in {cyan}) and a southern cone denoted as region \textit{3C9-b} (outlined in red). We also see a large Southern cloud in \heii\,(outlined in purple) denoted as region \textit{3C9-c}. For 4C 05.84, we see the biconical shape in \civ\, emission showing a {Northeastern} cone denoted as region \textit{4C0584-a} (outlined in {cyan}) and a {southwestern} cone (outlined in red) denoted as region \textit{4C0584-b}. We don't detect significant \heii\, emission in this region. We also see evidence of a larger {southeastern} cloud denoted as region \textit{4C0584-c} (outlined in purple) seen both in \civ\, and \heii. }%
    \label{fig:Q_Ratios_im}
\end{figure*}

\begin{figure*}[!ht]
    \centering
    {{\includegraphics[width=\linewidth]{ 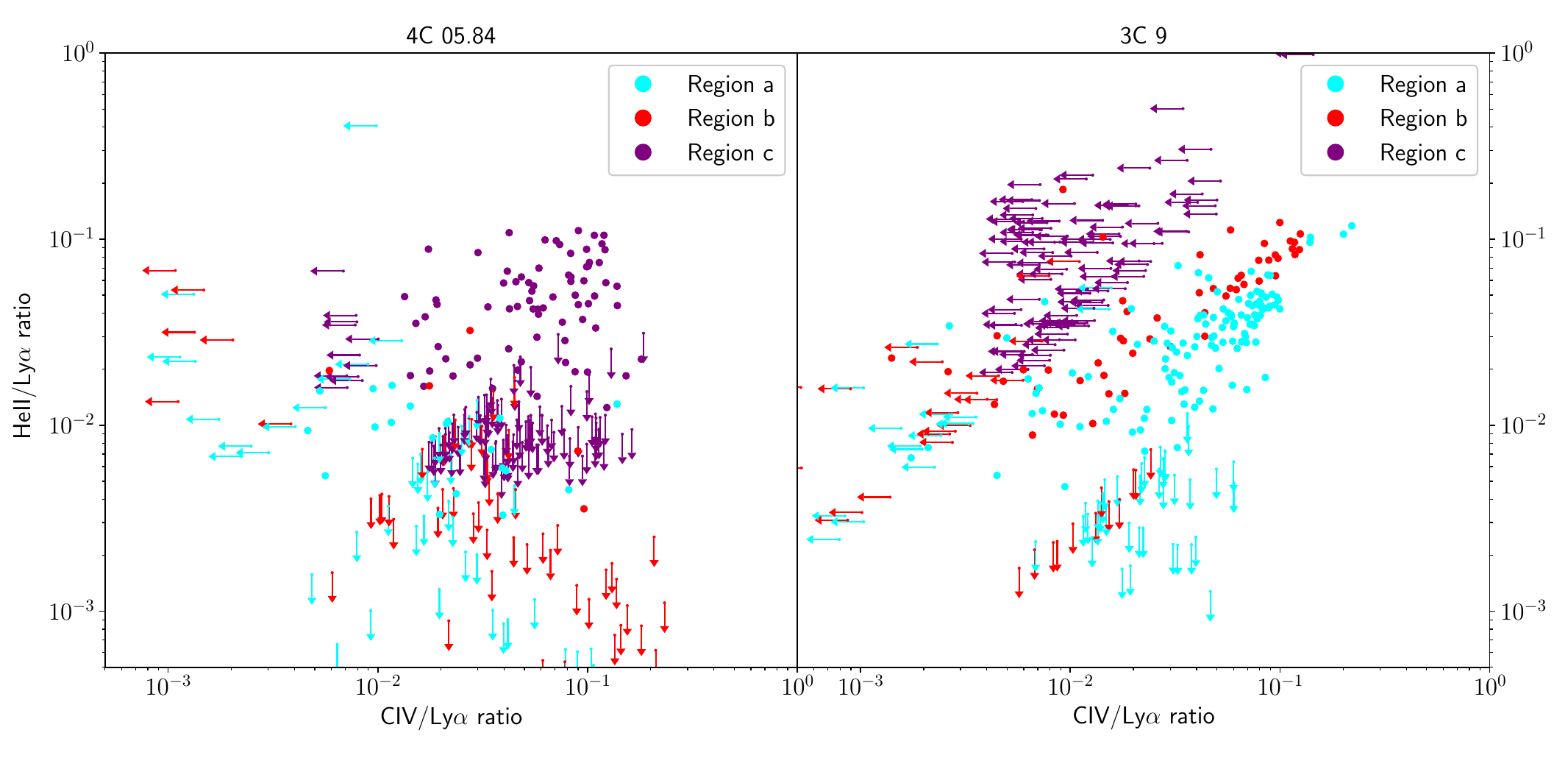} }}z
    \caption{We used the surface brightness ratio maps in Figure \ref{fig:Q_Ratios_im} and plot the \heiilya\, ratio against the \civlya\, ratio for individual spaxels for 4C 05.84 (Left) and 3C 9 (Right). We further distinguish pixels belonging to the three different regions of the CGM as discussed in text for each quasar. For 4C 05.84, we have region \textit{4C0584-a} ({Cyan}), region \textit{4C0584-b} (Red) and the region \textit{4C0584-c} (Purple) for 4C 05.84. Similarly, we have region \textit{3C9-a} ({Cyan}), region \textit{3C9-b} (Red) and region \textit{3C9-c} (Purple) for 3C 9. }%
    \label{fig:Q_Ratios_points}
\end{figure*}
{Previous works in the literature have indicated that \civlya\, and \heiilya\, ratios can be used to probe the metallicity and ionization in the CGM around a quasar \citep{Feltre2016, Guo2020, Lau2022}.} In this section, we use the surface brightness maps in Figures \ref{fig:KCWI-3C9} and \ref{fig:KCWI-4C0584} to construct spatial maps for \civlya\, and \heiilya. {We use these maps to explore the ionizing source and metallicity of the CGM around these two quasars in Section \ref{subsec:CGM_ionization_metallicity}.} These maps are presented in Figure \ref{fig:Q_Ratios_im}. Both quasars show biconical morphologies in \civ\, emission immediately around the quasar, and significant \heii\,emission around 20 kpc from the quasar. However, while we see \civ\, emission far away from the quasar for 4C 05.84 in \textit{4C0584-c}, this is not the case for 3C 9. A comparison with the moment 1 maps indicates that in both quasars, the surface brightness ratio is higher in the region with the relatively blueshifted velocity. Both the \textit{4C0584-c} and \textit{3C9-a} show a slightly larger extent for \civ\, emission than \heii. \\
Using these maps, we construct a \heiilya\, ratio vs. \civlya\, ratio plot for each spaxel, as shown in Figure \ref{fig:Q_Ratios_points}. We indicate points where no \heii\, or \civ\, is detected with limiting arrows. {Limits are calculated using 1$\sigma$ errors obtained from the Moment 0 error maps obtained from CWITools.} Based on the morphology and kinematics of the two quasars, we identify three distinct components defined previously in this plot, the regions \textit{3C9-a}, \textit{3C9-b}, \textit{3C9-c}, \textit{4C0584-a}, \textit{4C0584-b} and \textit{4C0584-c}. \heii\, emission in region \textit{3C9-c} shows no counterpart in \civ. We observe a clear trend in 3C 9 with higher \heiilya\, ratio seen in \textit{3C9-b}.

\subsection{Asymmetry of \lya\, halos}
\begin{figure*}
    \centering
    {\includegraphics[width=\linewidth]{ 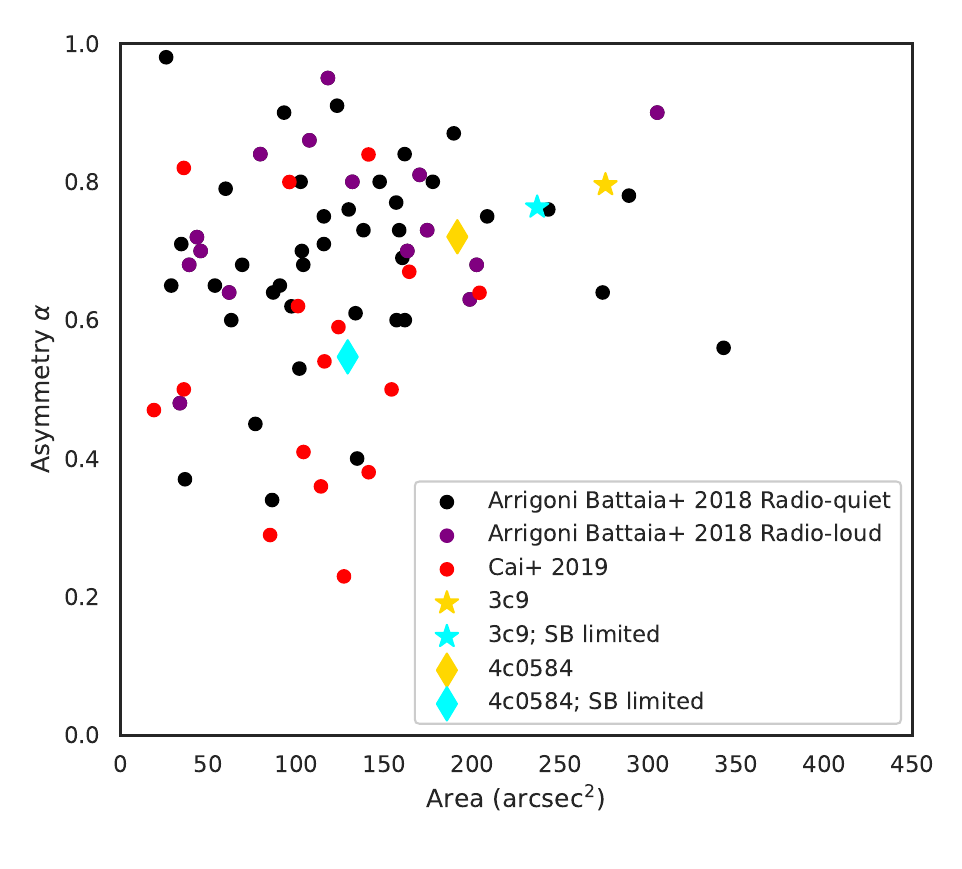} }%
    \caption{We plot the asymmetry $\alpha$ against the total area of the $2\sigma$ isophote for the two quasars 3C 9 and 4C 05.84. For comparison we plot these against the radio-loud and radio-quiet quasars from \cite{AB2019} at $z \sim 3$ and from the $z \sim 2$ quasars in \cite{Cai2019}. {The cyan markers for 3C 9 and 4C 05.84 represent datapoints if we only consider pixels with a minimum surface brightness limit of $2 \times 10^{-18}$ \fergarc. }}%
    \label{fig:asymmetry}
\end{figure*}

\begin{figure*}
    \centering
    {\includegraphics[width=\linewidth]{ 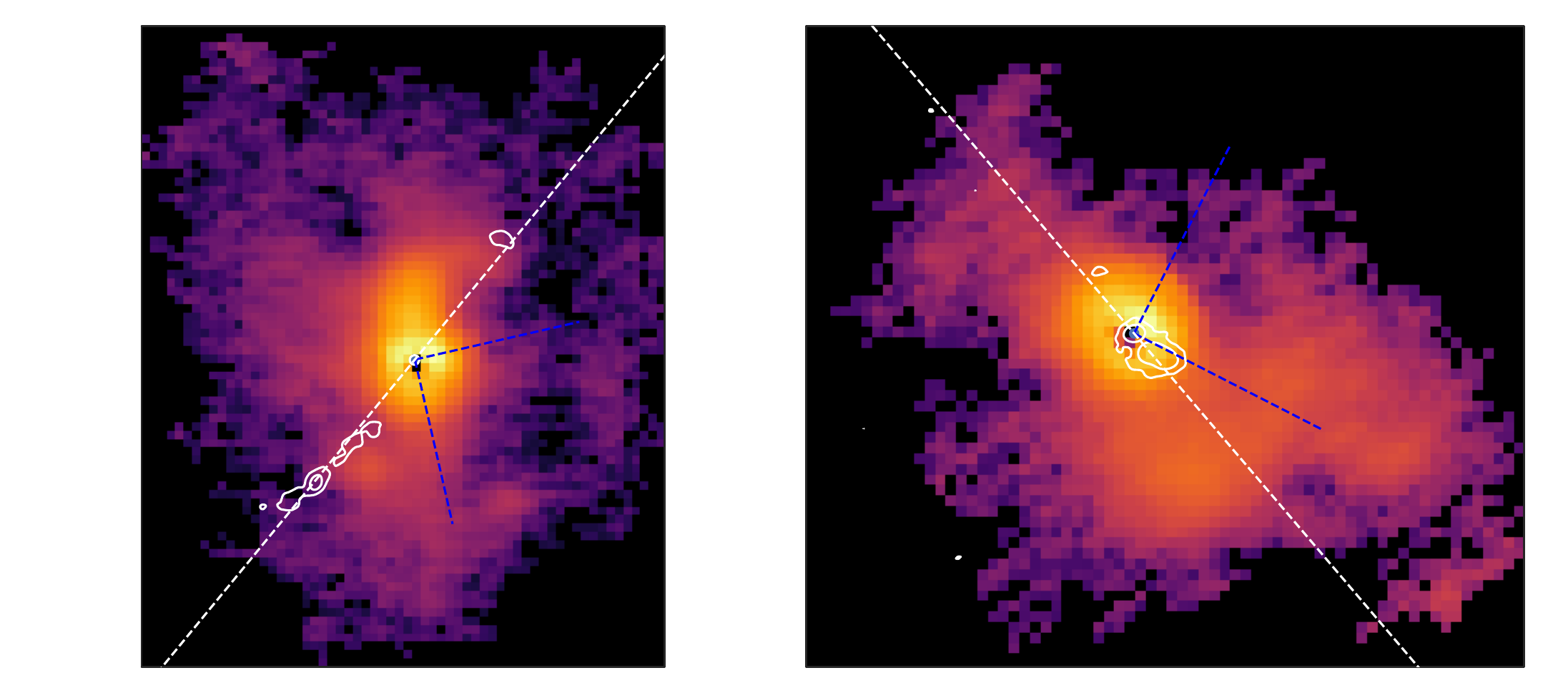} }%
    \caption{We use the angle $\phi$ determined using equation \ref{eq:phi} and determine the major and minor axes (blue dashed line) for the \lya\,nebula for 3C 9 (left) and 4C 05.84 (right). Using the right hand rule, the major axis is the X axis and the minor axis is the Y axis here. We also plot the radio jet contours and the jet axis and plot it on the same image (white dashed line). }%
    \label{fig:asymmetry_axes}
\end{figure*}

We calculate the asymmetry of the \lya\,nebulae by calculating the second order moment of the flux \citep{Stoughton2002}. These moments are defined as 
\begin{equation}
    M_{xx} = \langle \frac{(x-x_{Neb})^2}{r^2} \rangle_f ; \,
    M_{yy} = \langle \frac{(y-y_{Neb})^2}{r^2} \rangle_f 
\end{equation}

\begin{equation}
        M_{xy} = \langle \frac{(x-x_{Neb})(y-y_{Neb})}{r^2} \rangle_f
\end{equation}

Where $x_{Neb}$ and $y_{Neb}$ are flux weighted centroids for the $2\sigma$ isophotes of the nebulae and $r$ is the distance of a given pixel from the centroid. We can then calculate the Stokes parameters using the moments 
\begin{equation}
    Q = M_{xx} - M_{yy} ;
    U = 2 M_{xy}
\end{equation}
and derive the ratio between the minor and major axes of the nebula as the asymmetry $\alpha$ and the angle $\phi$ by using 
\begin{equation}
    \alpha = \frac{b}{a} = \frac{1 - \sqrt{Q^2 + U^2}}{1 + \sqrt{Q^2 + U^2}} 
\end{equation}
\begin{equation}\label{eq:phi}
    \phi = arctan(\frac{U}{Q})
\end{equation}
Here, the angle $\phi$ is the angle between the major axis and the nearest X or Y axis. The results are shown in Figure \ref{fig:asymmetry}. We also plot the major axis for the \lya\, nebulae and the radio jets in Figure \ref{fig:asymmetry_axes} We show the results from the $z \sim 3$ radio-loud and radio-quiet quasars from \cite{AB2019} and $z \sim 2$ quasars from \cite{Cai2019}. To provide a comparison at a similar surface brightness limit to previous studies (i.e., 1$\sigma$ limit in \cite{Cai2019} is $9 \times 10^{-19}$ \fergarc\, and 2$\sigma$ in \cite{AB2019} is $4 \times 10^{-18}$ \fergarc) we provide an asymmetry figure where we have considered a surface brightness limit of $2 \times 10^{-18}$ \fergarc\. The median projected areas at $z \sim 2$ and $z \sim 3$ are 63 $\rm{arcsec}^2$ and 123 $\rm{arcsec}^2$ respectively whereas the median asymmetries at these redshifts are 0.71 and 0.54 respectively. Once we take observational sensitivities into account, we find that 3C 9 is larger than the median \lya\, nebula at $z \sim 2$ and $z \sim 3$, but 4C 05.84 is very similar in area to nebulae at $z \sim 3$ but significantly larger than median area at $z \sim 2$. While 3C 9 is more symmetric than the $z=2$ \lya\, 4C 05.84 is again comparable to median asymmetry of $z \sim 2$ nebulae. While longer exposure times increased the detection area only modestly for 3C 9, with no notable change in the calculated asymmetry $\alpha$ value, the changes for 4C 05.84 are statistically significant. We conclude that 4C 05.84 is largely consistent with the expected asymmetry for nebulae at this redshift, whereas 3C 9 is more axi-symmetric than other samples. Both nebulae are larger than expected for $z \sim 2$ sources projected on-sky area. \\
We also determine the axis of the radio jet and compare this angle with that of the \lya\,nebula. We measure the angle between the radio and CGM axes to be 52.3 degrees for 3C 9 and 22.5 degrees for 4C 05.84.

\section{MOSFIRE Results: Group/cluster Member Properties } \label{sec:MOSFIRE}

\begin{deluxetable*}{rrrllcl}
\tablenum{2}
\caption{Redshift and Velocities of spectroscopically-confirmed group/cluster members  \label{tab:Group-members-coords}}

\tablehead{\colhead{Quasar System}&
\colhead{Source Name}&
\colhead{Redshift}&
\colhead{RA}&
\colhead{Dec}&
\colhead{Distance from Quasar}&
\colhead{Line Used}\\
\colhead{}&
\colhead{}&
\colhead{}&
\colhead{HMS}&
\colhead{DMS}&
\colhead{kpc}&
\colhead{}
}
\startdata
3C 9 & MOS-1 & 2.0221 $\pm$ 0.0001 & 00:20:25.198 & +15:40:57.602 & 24 & $[O III]$ \\
3C 9 & MOS-2 & 2.03002 $\pm$ 0.00004 & 00:20:25.244 & +15:40:49.409 & 44 & $H\alpha$ \\
3C 9 & MOS-3 & 2.0213 $\pm$ 0.0001 & 00:20:25.238 & +15:40:53.522 & 12 & $[O III]$  \\
3C 9 & MOS-4 & 2.0294 $\pm$ 0.0000 & 00:20:25.598 & +15:40:45.527 & 61 & $[O III]$  \\
3C 9 & MOS-5 & 2.02614 $\pm$ 0.00006 & 00:20:25.914 & +15:41:12.48 & 172 & $H\alpha$ \\
4C 05.84 & MOS-1 & 2.3239 $\pm$ 0.0001 & 22:25:14.568 & +05:27:05.673 & 32 & $H\alpha$ \\
4C 05.84 & MOS-2 & 2.3241 $\pm$ 0.0000 & 22:25:14.292 & +05:27:06.406 & 61 & $H\alpha$ \\
4C 05.84 & MOS-3 & 2.3246 $\pm$ 0.0000 & 22:25:14.336 & +05:27:01.162 & 38 & $[O III]$  
\enddata
\end{deluxetable*}
\begin{deluxetable*}{rrllllll}
\tablenum{3}
\caption{Line Fluxes in $10^{-17} \, erg \, cm^{-2} \, s^{-1}$ of spectroscopically-confirmed group/cluster members. \label{tab:Group-members}}
\tablehead{
\colhead{Source Name}&\colhead{\oiii}&\colhead{\hb}&\colhead{\ha}&\colhead{\nii}&\colhead{\lya}&\colhead{\heii}&\colhead{\civ}\\
\colhead{}&\colhead{}&\colhead{}&\colhead{}&\colhead{}&\colhead{}&\colhead{}&\colhead{}
}
\startdata
3C9-MOS-1 \tablenotemark{a} & $11.7 \pm 0.5$  & -\tablenotemark{b} & -\tablenotemark{d} & -\tablenotemark{d} & $369.1 \pm 0.01$ & $12.5 \pm 0.01$ & $1.66 \pm 0.002$ \\
3C9-MOS-2 & $18.3 \pm 0.3$ & $2.0 \pm 0.2$ & $5.8 \pm 0.3 $ & $0.32 \pm 0.06$ & $44.74 \pm 0.005$ & $1.73 \pm 0.01$ & -\tablenotemark{c} \\
3C9-MOS-3 & $ 47.3 \pm 0.3$ & $5.1 \pm 0.3 $ & $30.6 \pm 0.3$\tablenotemark{f} & -\tablenotemark{e} & -\tablenotemark{e} & -\tablenotemark{e} & -\tablenotemark{e}\\
3C9-MOS-4 & $7.5 \pm 0.3$ & $0.36 \pm 0.09$ & -\tablenotemark{e} & -\tablenotemark{e} & -\tablenotemark{e} & -\tablenotemark{e} & -\tablenotemark{e}\\
3C9-MOS-5 & -\tablenotemark{c} & $1.2 \pm 0.3$ & $5.8 \pm 0.2$ & $2.5 \pm 0.3$ & -\tablenotemark{d} & -\tablenotemark{d} & -\tablenotemark{d}\\
4C0584-MOS-1 & $15.0 \pm 0.3$ & $ < 0.4$\tablenotemark{c} & $2.7 \pm 0.25$ & -\tablenotemark{b} & $37.72 \pm 0.01$ & $2.7 \pm 0.005$ & $0.277 \pm 0.005$ \\
4C0584-MOS-2 & $0.78 \pm 0.09$ & $0.17 \pm 0.05$ & $0.75 \pm 0.2$  & $< 0.24$ & $19.09 \pm 0.01$ & -\tablenotemark{c} & -\tablenotemark{c}\\
4C0584-MOS-3 & $6.0 \pm 0.2$  & $< 0.26$\tablenotemark{c} & -\tablenotemark{d} & -\tablenotemark{d} & $23.89 \pm 0.02$ & $0.016 \pm 0.0005$ & $0.139 \pm 0.005$ \\
\enddata
\tablenotetext{a} {Redshift confirmed by using [O III 5007] and [O III 4959]}
\tablenotetext{b} {Emission line falls on top of a sky line and no flux could be extracted.}
\tablenotetext{c} {No flux detected.}
\tablenotetext{d} {No data available.}
\tablenotetext{e} {Flux Extraction affected by Quasar residuals.}
\tablenotetext{f} {{\ha\, flux extraction likely contaminated by \nii\, and must be treated as an upper limit.}}
\end{deluxetable*}
\begin{deluxetable}{cccc}
\tablenum{4}
\tablecaption{NIRC2 Magnitudes for identified sources \label{tab:NIRC2_mags}}
\tablehead{
\colhead{Source} & 
\colhead{J} & 
\colhead{H} & 
\colhead{Kp} \\ 
\colhead{} & 
\colhead{mag} & 
\colhead{mag} & 
\colhead{mag}
} 
\startdata
4C0584-MOS-1 & -\tablenotemark{a} & 22.8 & 22.0 \\
4C0584-MOS-2 & -\tablenotemark{a} & -\tablenotemark{a} & -\tablenotemark{a} \\
4C0584-MOS-3 & -\tablenotemark{a} & 24.6\tablenotemark{b} & 24.6\tablenotemark{b} \\
\enddata
\tablenotetext{a} {No source could be detected}
\tablenotetext{b} {Magnitude extracted as a 1.1$\sigma$ detection}
\end{deluxetable}
We use MOSFIRE to target potential group/cluster members around both quasars up to an angular separation of 3 arcminutes, which translates to a comoving distance of approximately 1.4 Mpc at the redshift of the quasars. The number of science targets we have for each mask are detailed in Table \ref{tab:Slitmasks}. At the quasar redshifts, we detect \ha\, in the K band and $H\beta$, $O[III] \, 5007 \AA$ and $O[III] \, 4959 \AA$ in the H band. Our target selection relies on a Spitzer color cut as described in Section \ref{sec:TarSel}. Table \ref{tab:Group-members-coords} lists spectroscopically identified MOSFIRE sources around the two quasars. We also list the extracted line fluxes for all spectroscopically detected members in Table \ref{tab:Group-members}. {Lastly, we show the extracted NIRC2 magnitudes for MOSFIRE confirmed sources in Table \ref{tab:NIRC2_mags}}.{We spectroscopically confirmed sources at $z > 1.3$ for 15 out of 62 targets in the 4C 05.84 field, and 19 out of 94 targets in the 3C 9 field. For identification of group/cluster members, we chose a redshift range of $z_{quasar} \pm 0.01$. We only confirm sources that have multiple emission lines detection, which leads to a lower confirmation rate. Given telescope time and weather conditions we were only able to complete K-band observations on only 1 mask for each quasar, to target \ha, which is expected to be prominent for these galaxies, can only be observed for a fraction of sources around each quasar. For 3C 9, \hb\, falls close to the edge of the atmospheric transmission, making it difficult to confirm any galaxies with faint \hb\, and \oiii\, emission at other redshifts. Identified redshifts, locations and properties of the extended dataset will be discussed in a future work.}

To calculate the redshift of the group/cluster members, we first visually inspect each source's spectrum and define a boxcar region to extract a 1D spectrum. We fit each spectrum with a Gaussian to determine the centroid and $\sigma$ for each emission line to determine the redshift $z_i$ of the member. For all cases, we have detection of multiple lines for the same source, and as such, the redshift is calculated using the line with the highest Signal-to-noise ratio. We then use the redshift of the quasar ($z=2.019$ for 3C 9 and $z=2.32$ for 4C 05.84 \citep{QUART_OSIRIS}) to determine the velocity offset of the companion source from the systemic redshift using $v_i = c \frac{z_i-z}{1+z}$. The fitted $\sigma$ gives us the resultant velocity dispersion of each source. 

We determine the astrometry for each MOSFIRE spectroscopically-confirmed source to extract a spatially-coincident spectrum in the KCWI data. We verify the MOSFIRE astrometry accuracy using the RA and Dec of the continuum source used in the science slit for each mask from the MOSFIRE pipeline header parameters. We additionally check for any optical distortions between the KCWI and MOSFIRE slitmasks by checking the pointing of the MOSFIRE slitmask on the quasar. We then calculated slit offsets for nearby objects from the quasar in MOSFIRE and KCWI WCS coordinates and found them to be in good agreement. We found that the distortion between the two instruments does not cause a significant offset in the astrometry {($<$0.1 arcsecond)}. We show the locations of the MOSFIRE slits indicating the spatial extent of detected emission lines in Figure \ref{fig:MOS_Slits}. 

\subsection{KCWI and MOSFIRE Spectra for all sources}
\begin{figure*}[ht!]
    \centering
    {{\includegraphics[width=\linewidth]{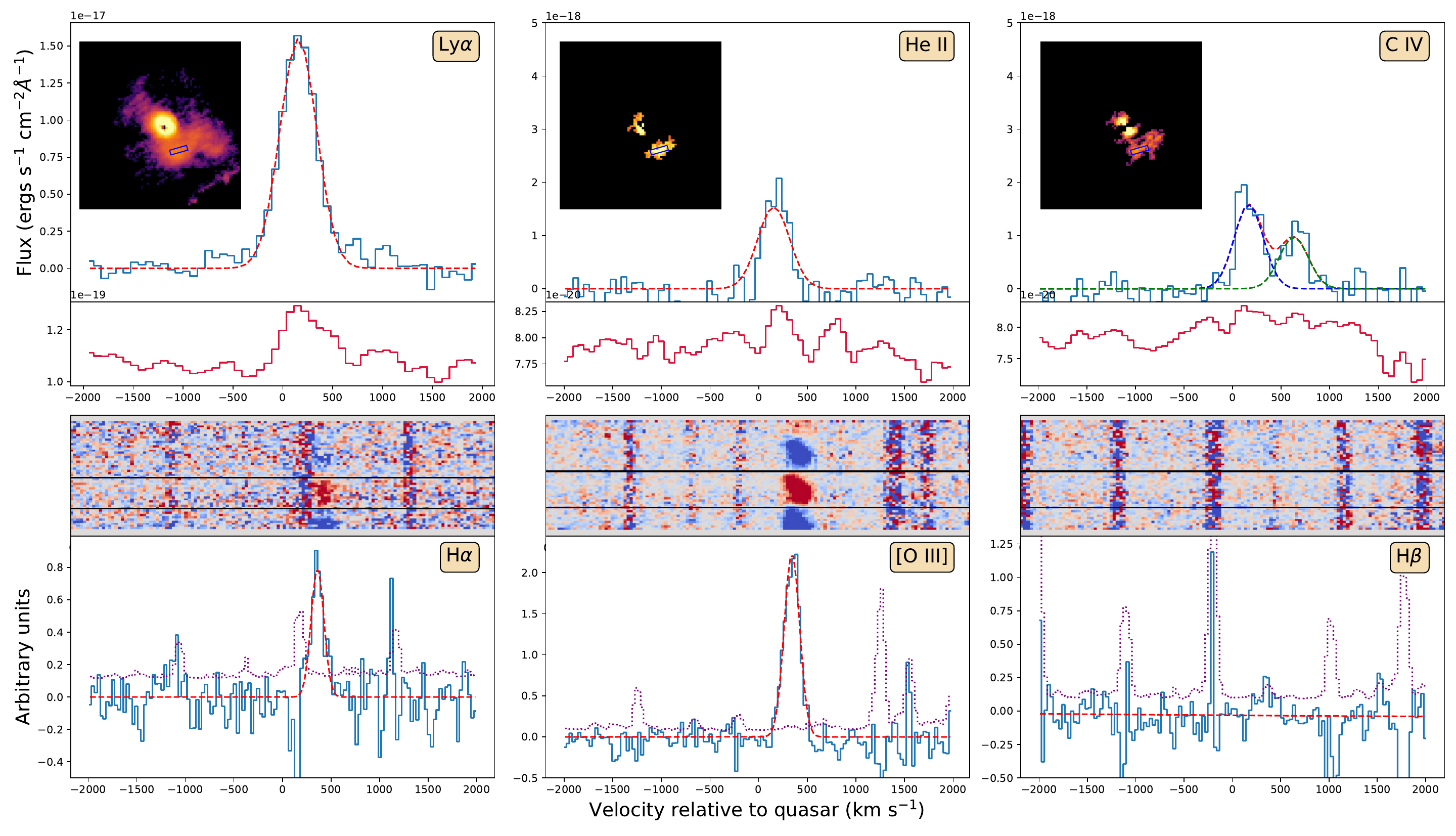} }}%
    \caption{Figure showing all spectra for the source 4C 05.84 - MOS-1 from KCWI and MOSFIRE data. {The X-axis for all spectra shows the velocity with respect to the quasar's redshift in \kms. For the top row, the Y-axis shows the extracted 1D spectrum in \ferg, whereas for the bottom row, the Y-axis is in relative units.} The top row contains 1D spectra extracted from KCWI Surface Brightness plots as shown in Figure \ref{fig:KCWI-4C0584}. The red dashed line indicates the best fit gaussian for each emission line whereas the propagated scaled uncertainties for the measurements are shown in the y axis split plot shown in red. Each emission line shows an inset of the KCWI Surface brightness image with a rectangular slit (blue) corresponding to MOSFIRE emission line detection showing the location where the KCWI 1D spectrum was extracted from. The bottom row consists of MOSFIRE detected \ha, \hb\, and \oiii\, emission lines with the 2D spectra on top and the extracted 1D spectrum between the horizontal boxcar region (shown with solid black horizontal lines in the 2D spectrum) on the bottom. The red dashed line indicates the best fit gaussian whereas the purple dotted line indicates the peak sky noise. The complete figure set (3 images) is available in the online journal.}%
    \label{fig:MOS-4C0584}
\end{figure*}

\begin{figure*}[ht!]
    \centering
    {{\includegraphics[width=\linewidth]{  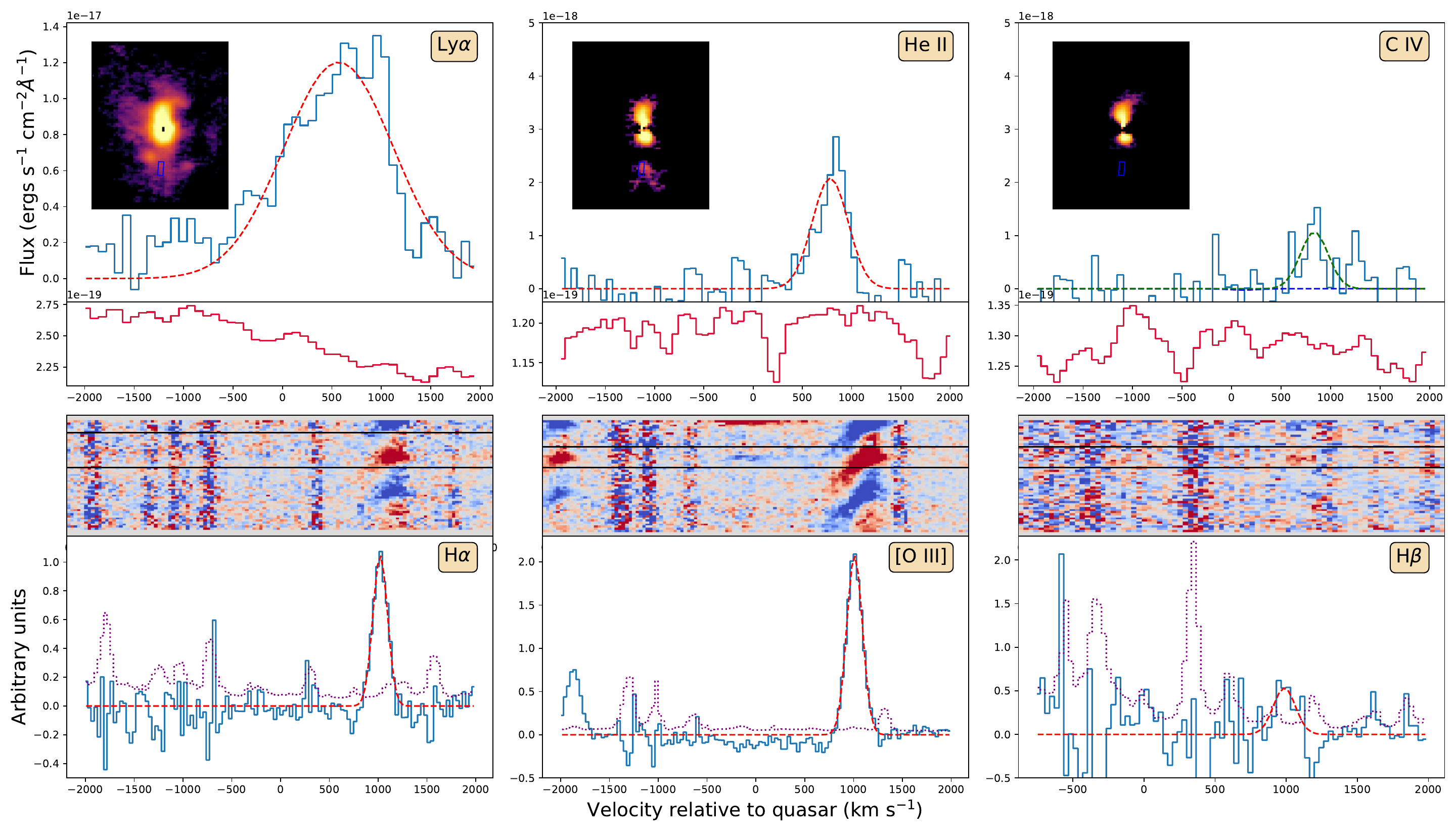} }}%
    \caption{Figure showing all spectra for the source 3C 9 - MOS-2 from KCWI and MOSFIRE data with all axes and labels similar to Figure \ref{fig:MOS-4C0584}. The complete figure set (5 images) is available in the online journal.}%
    \label{fig:MOS-3C9}
\end{figure*}

We plot all extracted 1D MOSFIRE spectra for all identified sources in this section. All the spectra are available in the online version of this journal for 4C 05.84 (3 Figures) and 3C 9 (5 Figures). We show all extracted spectra for 4C0584-MOS-1 in Figure \ref{fig:MOS-4C0584} and 3C9-MOS-2 in Figure \ref{fig:MOS-3C9} as examples for the two Figure Sets. Identical figures for all the other sources can be found in the Appendix. Wherever possible, we have fitted identified lines with a Gaussian function. In some cases, the best fit requires us to fit a broad and narrow Gaussian component. In the case of 3C9-MOS-3 and 3C9-MOS-4, we have added a continuum fit to take into account any flux contamination from the quasar. {Since \ha\, can contaminate flux extraction from \nii\, and vice versa, we always fit two gaussians when extracting flux. For most sources where \ha\, and \nii\, 6583.46 $\AA$ is observed, we observe no contamination as the two lines can be visually distinguished. The only exceptions are 3C9-MOS-3 and 3C9-MOS-4 spectra, where we do see a blending between the two lines. In these cases, we were unable to distinguish the fluxes from \ha\, and \nii\, emission lines. Thus, for 3C9-MOS-3, we considered the extracted flux for the \oiii\, and consider the same velocity and spatial extent to be true for \ha\, and report this as the \ha\, flux. We do not report any flux for \nii\, for this object. We are unable to apply the same process to 3C9-MOS-4, so no \ha\, or \nii\, flux is reported for this object. We note in Table \ref{tab:Group-members} that this \ha\, flux could be affected by \nii\, and must be treated as an upper limit.} \\
We overlay the MOSFIRE slit aperture on the KCWI emission line maps and extract a 1D spectrum for \lya, \heii\, and \civ\, corresponding to the location of the MOSFIRE slit for comparison. We plot 1 source here from each quasar for reference. The complete figure set (3 images for 4C 05.84 and 6 images for 3C 9) is available in the online journal.\\
We show in Table \ref{tab:CGM-Galaxy-sources} the spectroscopically identified MOSFIRE galaxies and the kinematically distinct regions of CGM they are associated with as seen with KCWI. This will be useful to keep in mind particularly when we discuss our results in section \ref{sec:Discussion}. Whenever we refer to the MOSFIRE identified galaxies and any properties associated with the ISM of that galaxy (e.g. star formation rate), we use the MOSFIRE identifier for that galaxy. Whenever we discuss the CGM properties of distinct regions around the quasar, we use the region identifiers established in Figure \ref{fig:Q_Ratios_im}.\\

\begin{deluxetable*}{cccc}
\tablenum{5}
\caption{CGM regions around the two quasars and the associated MOSFIRE galaxies \label{tab:CGM-Galaxy-sources}}
\tablehead{\colhead{Quasar System}&
\colhead{CGM Region}&
\colhead{MOSFIRE Identified Galaxies}&
\colhead{Transition used for Definition}
}
\startdata
3C 9 & 3C9-a & 3C9-MOS1, MOS 3C9-MOS-3 & \civ \\
3C 9 & 3C9-b & 3C9-MOS-4 & \heii \\
3C 9 & 3C9-c & 3C9-MOS-2 & \heii \\
4C 05.84 & 4C0584-a & - & \civ\\
4C 05.84 & 4C0584-b & - & \civ\\
4C 05.84 & 4C0584-c & 4C0584-MOS-1, 4C0584-MOS-3 & \civ
\enddata

\end{deluxetable*}

\subsection{Star Formation Rates \& Metallicities}

We convert \ha\, fluxes into an estimation of star formation rates (SFR) from \cite{SFR2011Murphy}. We use the expression -
\begin{equation}
\frac{\rm{SFR}_{H \alpha}}{M_{\odot} yr^{-1}} = 5.37 \, \times \, 10^{42} \, \frac{L_{H \alpha}}{\rm{ergs \, s}^{-1}}
\end{equation}
Wherever possible, we have corrected the \ha\, flux based on a calculation of the Balmer decrement. Using the attenuated to intrinsic line ratio $R_{\alpha \beta}$ assuming Case B Recombination at a temperature of 10,000 K, we calculate the value of $E(B-V)_{nebular}$ using results from \cite{Calzetti2001DustCorrection}. We then convert this into $E(B-V)_{continuum}$ using the conversion \citep{Calzetti2000EBV} - 
\begin{equation}
    \rm{E(B-V)}_{nebular} = (0.44 \pm 0.03) \times \rm{E(B-V)}_{continuum}
\end{equation}
We then follow \cite{Steidel2014KBSS} to assume either the Galactic or the Small Magellenic Cloud (SMC) Extinction curve depending on the measured E(B-V), to obtain the \ha\, attenuation in magnitudes as given by -
\begin{equation}
    A(H \alpha) = 4.54 \ \rm{E(B-V)}_{cont}; \ \rm{E(B-V)} \leq 0.20
\end{equation}
\begin{equation}
    A(H \alpha) = 5.72 \ \rm{E(B-V)}_{cont}; \ \rm{E(B-V)} > 0.20
\end{equation}
For sources where we do not have \ha\, data but only \hb, we assume Case B Recombination nebular conditions with T=10,000 K giving an \ha/\hb\, ratio of 2.86 to calculate \ha\, flux, which we then use to compute the SFR for each source. Since the \hb\, line falls particularly close to low atmospheric transmission window for 3C 9 and close to sky lines in the H band for both quasar, we find that the calculated \ha/\hb\, ratios can be as high as $\sim 6$. These should not be taken to imply dust extinction but are likely due to a combination of relatively low SNR for the \hb\, line and the difficulty in carrying out accurate telluric calibration on the edge of the atmospheric transmission window.\\
For each source we calculate metallicities using O3N2 = log((\oiii/\hb)/(\nii $\lambda$6584/\ha)) and N2 = log(\nii$\lambda$6583/\ha) to calculate metallicites where data is available using calibration from \citet{PP04}. In cases where we do not detect \nii at a 2$\sigma$ limit, we have instead provided a limit. For objects where we do not have any \ha\, and \nii $\lambda$6584 data, we instead use results from \citet{NMM06} to provide an upper limit on the 12 + log(O/H) values using \oiii/\hb. {We estimate virial masses for detected MOSFIRE sources that are spatially resolved. We choose the extent of the object as detected with MOSFIRE to be the total radius and the velocity dispersion measured by MOSFIRE to be the line-of-sight velocity. We then calculate the total mass by converting the line-of-sight velocity into 3d velocity for the source, assuming an isotropic velocity distribution.} The extracted metallicities, SFRs and an estimation of the virial mass for each source is listed in Table \ref{tab:Member_props}.

\begin{deluxetable*}{lcccccc}
\tablenum{6}
\tablecaption{Group/cluster Member properties \label{tab:Member_props}}
\tablehead{
\colhead{Source} & 
\colhead{12 + log(O/H)} & 
\colhead{12 + log(O/H)} & 
\colhead{12 + log(O/H)} & 
\colhead{$\rm{SFR}_{H\alpha}$} & 
\colhead{Virial Mass} & 
\colhead{Line Used}  \\ 
\colhead{} & 
\colhead{(O3H$\beta$)} &
\colhead{(N2)} &
\colhead{(O3N2)} &
\colhead{\myr} & 
\colhead{$\times 10^{10}$\msun} & 
\colhead{}
} 
\startdata
3C9-MOS-1 & - & - & - & 19.8\tablenotemark{a} & 16.7 & \oiii\, \\
3C9-MOS-2 & $< 8.42$ & 8.47 & 8.34 & 10.0 & 1.2 & \oiii\, \\
3C9-MOS-3 & - & - & - & 257 & - & \oiii \\
3C9-MOS-4 & - & - & - & 1.7\tablenotemark{b} & 3.4 & \oiii\, \\
3C9-MOS-5 & $< 9.3$ & 8.80 & - & 30.3 & 5.7 & \ha\, \\
4C0584-MOS-1 & - & - & - & 6.4 & 1.6 & \ha\, \\
4C0584-MOS-2 & $< 8.64$ & 8.61 & 8.50 & 3.73 & 0.56 & \ha\, \\
4C0584-MOS-3 & - & - & - & 14.2\tablenotemark{a} & 2.3 & \oiii\, \\
\enddata
\tablenotetext{a} {Calculated assuming \ha\, flux equal to the \oiii\, flux.}
\tablenotetext{b} {Calculated using \hb\, and assuming Case B Nebular conditions.}

\end{deluxetable*}

\subsection{Nebular Diagnostic Diagrams}

\begin{figure*}
    \centering
    {{\includegraphics[width=\linewidth]{ 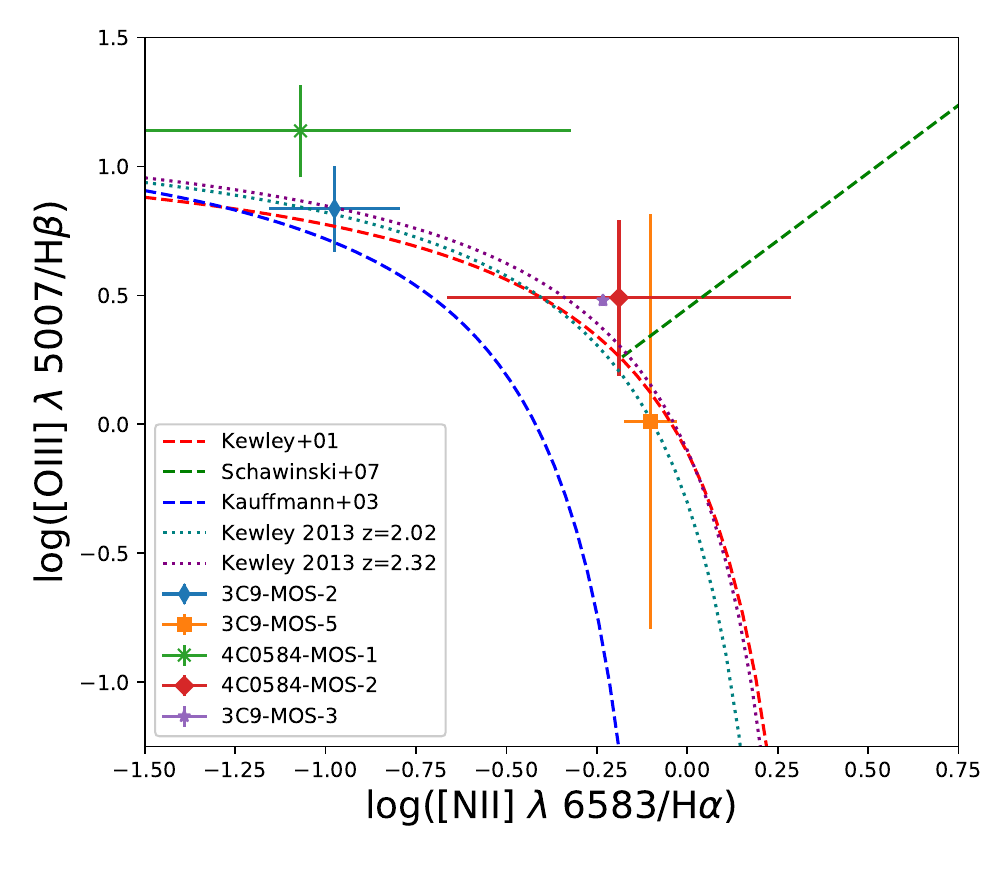} }}%
    \caption{Diagnostic diagram (log(\oiii/\hb) vs. log(\nii/\ha))\citep{BPT} for all sources detected at the redshift of the respective quasars using MOSFIRE. The star forming sequences from \cite{Kewley2001} and \cite{Kauffmann2003} are plotted to demarcate the star forming sequence. We include the sequence with redshift dependence for the two quasars as taken from \cite{Kewley2013}, including the line demarcating AGNs from LINERs as taken from \cite{Schawinski2007}. }%
    \label{fig:bpt_diagram}
\end{figure*}

We construct Baldwin, Phillips and {Terlevich} (BPT) diagrams \citep{BPT} for the companion galaxies using MOSFIRE fluxes. We plot these values against star forming sequences indicated in \citet{Kewley2001} and \citet{Kauffmann2003}, and the Seyfert-LINER separation as defined in \citet{Schawinski2007}. We include the results from \citet{Kewley2013} for the evolution of the star forming sequence as a function of redshift, highlighting the sequences at the redshifts of our quasars.\\

To calculate the emission line ratios, we extract flux for an aperture size that fits all four emission lines spatially and spectrally with flux greater than 2$\sigma$. We then propagate measurement uncertainties in our calculation of the ratio. The results are plotted in Figure \ref{fig:bpt_diagram}. \\
We had sufficient spectral coverage to investigate line ratios for 5 MOSFIRE sources. We now address all 5 sources individually. Firstly, we notice that 3C9-MOS-3, which is the quasar host galaxy, is the only source where the observational uncertainties are small enough that we can definitively categorize it as an AGN. This is not surprising. However, all other sources have significant uncertainties in the calculated line ratios. \\
Firstly, we consider the sources around 3C9, i.e. 3C9-MOS-2 and 3C9-MOS-5. Both of these sources lie on the star forming sequence. The question to consider is whether these can be ionized by an AGN given the large uncertainties. For 3C9-MOS-2, which is closer to the quasar, this is certainly a possibility. However, given that we see the stellar continuum from the galaxy in HST imaging, it is equally plausible that this galaxy is indeed on the star forming sequence. The same argument is even stronger for 3C9-MOS-5, as it is significantly further away from the quasar and detected independently with Spitzer and HST. We conclude that these galaxies are likely on the star forming sequence.\\
For 4C0584, the picture is less clear. We detect a strong continuum in NIRC2 imaging for 4C0584-MOS-1 but a much weaker component for 4C0584-MOS-2. While it does appear that both these sources are likely ionized by an AGN, there are important considerations here. Firstly, for 4C0584-MOS-1, the strong stellar continuum detection implies that it is possible that there is at least a combination of AGN and star forming activity that is causing the ionization of the galaxy. This seems less likely for 4C0584-MOS-2, which on sky appears to be further away from the quasar than 4C0584-MOS-1, but this could be due to on sky projection. The other complicating factor is that for 4C0584-MOS-1, the \nii\, line flux is contaminated by a sky line. This explains the particularly large uncertainty in the \nii/\ha ratio calculation which would bring it in line with 4C0584-MOS-2, if the \nii\, flux was strong. Given the emission line fluxes and their uncertainties, it seems likely that at least some part of the emission is powered by AGN photoionization. Since our calculated star formation rates for these galaxies have been derived using the nebular line fluxes, these rates should be treated as upper limits for these galaxies, as the true nebular emission due to star formation is a fraction of the observed flux. However, this interpretation does not affect our mass calculation of the system, which only relies on the observed positions and velocities of the companion galaxies.

\subsection{Calculating the Dark Matter mass of the two systems}\label{sec:MassCalc}
{The identification of companion galaxies allows us to calculate the dynamical mass of the dark matter halo. There are different approaches in the literature that calculate dark matter halo masses for a group/cluster of galaxies at comparable redshifts. Firstly, clustering strengths of galaxies in a group/cluster environment can be compared to simulated population of dark matter halos to estimate the halo mass. This is usually carried out by calculating galaxy-galaxy autocorrelation lengths, given a group of galaxies \citep{Trainor2012, White2012}. The drawback to this approach is that the robustness of the autocorrelation length calculation improves with the number of galaxies, and thus this method yields reliable results only with large number of galaxies. Secondly, one can use halo mass ($M_{DM}$) - stellar mass ($M_{stellar}$) relations to calculate the DM halo mass \citep{Moster2010,Moster2018}. This method has been validated by comparison with characteristic group luminosities using SDSS \citep{Yang2007} and has subsequently used to calculate group masses at lower redshifts \citep{Fossati2019}. However, this method relies on accurate SED modelling, thus works best when the galaxies in question have imaging across a wide range of optical and infrared bands. Other works have utilized this technique to estimate DM halo masses at higher redshifts. \cite{AB2022} use this method to estimate the DM halo mass of a z $\sim \, 3$ QSO environment and find a halo mass of $8.6 \times 10^{12} $ M$_{\odot}$, the bulk of which comes from the QSO host galaxy DM mass. Similarly, \cite{Chen2021} find a DM halo mass of $6.4 \times 10^{12} $ M$_{\odot}$ for the Slug nebula around the radio quiet quasar UM 287 at $z=2.23$. In both of these cases, the authors recognize the uncertainty in the calculations come largely from the estimation of the stellar masses of the galaxies in the environment.}\\

In this work, we follow the procedure laid out by \citet{Tempel2014} to calculate the dynamical mass of the environments. {\cite{Chen2021} and \cite{AB2022} use this as an alternate method for their DM mass calculations and find good agreement with the estimates from M$_{DM}$ - M$_{stellar}$ relations.} Given the velocity dispersion along line of sight to be $\sigma_{LOS}$ and assuming a virialized system, the three dimensional velocity dispersion is given by $\sigma_{3D} = \sqrt{3} \, \sigma_{LOS}$. Then, assuming an NFW profile \citep{NFWProfile} for the dark matter halo, we can iteratively calculate the projected radius on sky and the corresponding $M_{200}$. Having calculated the projected radius, the enclosed mass is given by -
\begin{equation}\label{eq:mass}
    M_{tot} = 2.325 \times 10^{12} \, \frac{R_g}{Mpc} \, \left(\frac{\sigma_{3D}}{100 \, km \, s^{-1}} \right)^2 \, M_{\odot}
\end{equation}
We obtain a projected radius of 123 kpc for 4C 05.84 using this method and a velocity dispersion of 352 \kms\, which gives a total mass of $3.5\times 10^{12} M_{\odot}$. For 3C 9, we get a projected radius of 263 kpc and a velocity dispersion of 673 \kms\, which gives us a total mass of $2.77\times 10^{13}\, M_{\odot}$.\\
We now address a concern regarding the mass calculation for 4C 05.84. For both quasars, we have calculated the mass of the dark matter halos under the assumption that the emission lines traced by MOSFIRE are tracing emission from companion galaxies and not simply emission from the CGM or any gas blobs in the vicinity of the quasar host galaxies. We believe we are indeed tracing the velocities and locations of galaxies since these sources were selected based on imaging of their stellar continuum using either HST F160W or as seen with Spitzer. However, for 4C0584-MOS-3, while the source was identified on the basis of continuum emission with Spitzer IRAC imaging, the source was close enough to the quasar such that the PSF from the quasar blended with the source. NIRC2 imaging of the quasar does not show a stellar continuum at the location of 4C0584-MOS-3. While it is possible that this is due to the sensitivity of the NIRC2 exposures, we consider the consequences if 4C0584-MOS-3 is indeed tracing the CGM and not stellar emission from a companion galaxy. When we remove 4C0584-MOS-3 from our mass calculation, we calculate the dark matter halo mass of the system to be $5 \times 10^{12} M_{\odot}$, larger than our original estimate. This is because MOS-3 does not serve as an outlier in physical separation from the rest of the group members, and has a velocity relatively close to the mean velocity of the system. As a result, when excluded from the calculation, it does not greatly affect the extent of the dark matter virial radius, but the expected velocity dispersion does increase, resulting in a net increase dark matter halo mass. However, since a virial mass calculation of the source 4C0584-MOS-3 indicates a mass $\sim 10^{10} M_{\odot}$, it is reasonable to assume this compact area could originate from star formation within a galaxy. Thus, for purposes of estimation of growth of these dark matter halos as discussed in Section \ref{subsec:Mass_growth}, we assume the originally calculated mass as a conservative estimate with the caveat that the calculated mass growth must be seen as a lower limit.

\section{Discussion} \label{sec:Discussion}

\subsection{CGM and companion galaxies}
\lya\, is a resonant emission line with a large optical depth. Thus the observed spatial profile does not necessarily coincide with the ionizing region. \heii\, emission, on the other hand, arises from a much more localized environment. We detect multiple \heii\, and \civ\, clouds with either a high-velocity offset or a large radial separation from the quasar --- or both --- using KCWI, which indicates the presence of other galaxies in the environment. With MOSFIRE follow up we confirmed the presence of the companion galaxies by identifying \ha, \hb\, and \oiii\, emission lines. Knowing the locations of the companion galaxies, we can also confirm that the peak in surface brightness in \heii\, away from the quasar host galaxy as seen in Figure \ref{fig:Q_Radial_Profile} for both 3C 9 and 4C 05.84 in \heii\,originates from companion galaxies. The case of \lya\, however, is more complicated. While we do detect a peak in \lya\, emission indicative of substructure at the radial distance of companion galaxies around 3C 9 and 4C 05.84, in the case of 3C 9 at least, it is clear from Figure \ref{fig:KCWI-3C9} surface brightness profile that the \lya\, emission peak does not necessarily coincide with the emission from the companion galaxies. In fact, by plotting the \lya\, emission, locations of companion galaxies and the locations of radio jets together as seen in Figure \ref{fig:MOS_Slits}, we see that for 4C 05.84, the \lya\, surface brightness peak in the region \textit{4C0584-c} happens to coincide with the direction of the radio jet and the companion galaxies. However, for 3C 9, the \lya\, emission seems to be coincident with the radio jets rather than companion galaxies. \\

For 4C 05.84, the \heii\, emission in \textit{4C0584-c} coincides with \civ\, emission as well as \ha, \hb\, and \oiii\, emission observed using MOSFIRE. The large spatially resolved extent of the \heii\, and \civ\, in \textit{4C0584-c} indicates that these are part of the companion galaxies 4C0584-MOS-1 and 4C0584-MOS-2.\\

Regarding 3C 9, we observed \heii\, emission away from the quasar in region \textit{3C9-c} but no spatially resolved \civ\, emission in that region. But we detect \ha, \hb\, and \oiii\, coincident with the \heii\, emission. If we assume that the quasar is responsible for photoionizing regions \textit{3C9-a} and \textit{3C9-b}, this implies the ionizing luminosity must be sufficient to cause \civ\, emission in \textit{3C9-c}. The absence of \civ\, emission thus indicates the presence of lower gas metallicity in region \textit{3C9-c}. This finding is supported by Figure \ref{fig:Q_Ratios_points}, which shows the \heiilya\, ratio against the \civlya\, ratio for all spaxels in the KCWI field of view. The figure demonstrates that the absence of \civ\, detection in \textit{3C9-c} is not due to observational sensitivity. This is particularly interesting as this is consistent with the observed \nii/\ha\, ratio for the galaxy which indicates a low metallicity. 

\subsection{Proto group/cluster mass growth} \label{subsec:Mass_growth}
To estimate the growth of these dark matter halos, we calculate the growth rate using two independent approaches. First, we use analytic relations derived using the extended Press-Schechter formalism to calculate the growth rate of the dark matter halos \citep{Correa2015a, Correa2015b, Correa2015c}. Second, we compare the dark matter masses to the merger trees constructed using the joint set from Millenium and Millenium-II simulations \citep{Fakhouri2010}.\\
Using the analytic Press-Schechter formalism, we use WMAP9 cosmological parameters to estimate the mass growth of the halos. We find that 4C 05.84 is accreting mass at a rate of approximately $2.3 \times 10^3$ \myr at $z=2.32$ and will grow to a mass of $2.2 \times 10^{13}$ \msun by present day. The same approach gives a mass accretion rate of $2 \times 10^4$ \myr for 3C 9 at $z=2.02$. We calculate 3C 9 will grow to a mass of $2.35 \times 10^{14}$ \msun by present day. \\

We compare these results to the halo merger trees constructed using the joint data set from the Millenium and Millenium-II simulations. \cite{Fakhouri2010} found an analytical fit to the mass accretion rate for the simulated halo merger trees given by -

\begin{equation}
\begin{split}
    \left<\dot{M}\right>_{mean} = & \, 46.1 \ \mathrm{ M}_{\odot} \, \mathrm{ yr}^{-1} \left( \frac{M}{10^{12} \, \mathrm{M}_{\odot}} \right)^{1.1} \\
    & \times \left( 1 + 1.11z \right) \sqrt{\Omega_m \left(1+z^3 \right) + \Omega_{\Lambda}}
\end{split}
\end{equation}

Starting with a range of DM halo masses at present-day, we calculate the progenitor mass by integrating the mass accretion rate as a function of redshift. We find the DM halo mass at present-day matches the dynamical mass of the quasar systems as calculated in Section \ref{sec:MassCalc} at the redshift of the quasar. Using this approach, we calculate that 4C 05.84 will grow to a mass of $2.3 \times 10^{13} \, \mathrm{M}_{\odot}$ and 3C 9 will grow to a mass of $2.15 \times 10^{14} \, \mathrm{M}_{\odot}$ by present-day, in reasonable agreement with our extended Press-Schechter calculation. This implies that the 4C 05.84 system will grow to a mass 10 times more massive than the local group \citep{Benisty2022}, making it a large group. Whereas 3C 9 will grow to a mass comparable to the dynamical mass estimates of Abell 2495 ($2.8 \pm 0.7 \times 10^{14}$ \msun) or MKW3S ($2.3 \pm 0.6 \times 10^{14}$ \msun) clusters \citep{Sifon2015}. This implies the 3C 9 system is in fact a protocluster environment. \\
\cite{Fakhouri2010} also calculate and fit the mean number of mergers a DM halo of mass M is expected to undergo between redshift of z and present day, with progenitor mass ratios in the range of $\xi = 0.1$ and $\xi = 0.3$. Based on their results, DM halos with a mass of 4C 05.84 had an average of 0.9 mergers between $z=2.32$ and today, whereas for 3C 9 it was an average of 1 merger between $z=2.02$ and now. Based on the virial mass estimates for the group/cluster members as listed in Table \ref{tab:Member_props}, all group/cluster members detected fall within the expected mass range to be merger candidates. If we treat the relative velocity between group/cluster members and the quasar as a velocity of approach, we can calculate the dynamical time for the galaxies to merge. The maximum time is $\sim$140 Myrs for the 3C9-MOS-5 source to merge with 3C 9{, assuming the true separation is equal to the projected on-sky separation. This is likely an underestimate of the distance and therefore the merger may occur later. However, even if the projected distance is underestimated by a factor of 10, all the companion sources detected have sufficient time compared to estimated dynamical time to merge with the quasar host.} The number of mergers thus would be 2-3 for each quasar system, which is higher than the expected number of mergers seen with the Millenium halo merger trees. {While these two estimates differ by a factor of 2 or 3}, the simulations usually calculate these as an average across a large number of merging systems. Given the stochastic nature of individual merger systems combined with small number statistics means we do not consider this result significantly different from the simulations.\\

The mass growth estimates calculated here rely on the initial mass of the DM halo, which has been calculated using the observed companion galaxies' positions and relative velocities with respect to the quasar. However, since it is possible there are galaxies further away from the quasar which were not detected, the calculated dynamical mass should be seen as a lower limit, which implies the same is true for the calculated mass accretion rates and the present day masses of these systems. We would like to stress that these calculations are approximations and can change based on new or improved data and that these masses must be treated as lower limits. We also caution against a simple interpretation of these structures as a protogroup or protocluster. We have chosen to demarcate the definition of a cluster as a bound structure that evolves to $z=0$ with a minimum mass of $10^{14}$ \msun, consistent with literature \citep{Overzier2016}. However, galactic structures lie on a mass spectrum and we would like to emphasize that with the aforementioned definition of a cluster and given our calculations, 4C 05.84 would be a large group whereas 3C 9 will be a small cluster by $z=0$.

\subsection{Inner and Outer CGM component model}
\begin{figure*}[!ht]
    \centering
    {{\includegraphics[width=\linewidth]{ 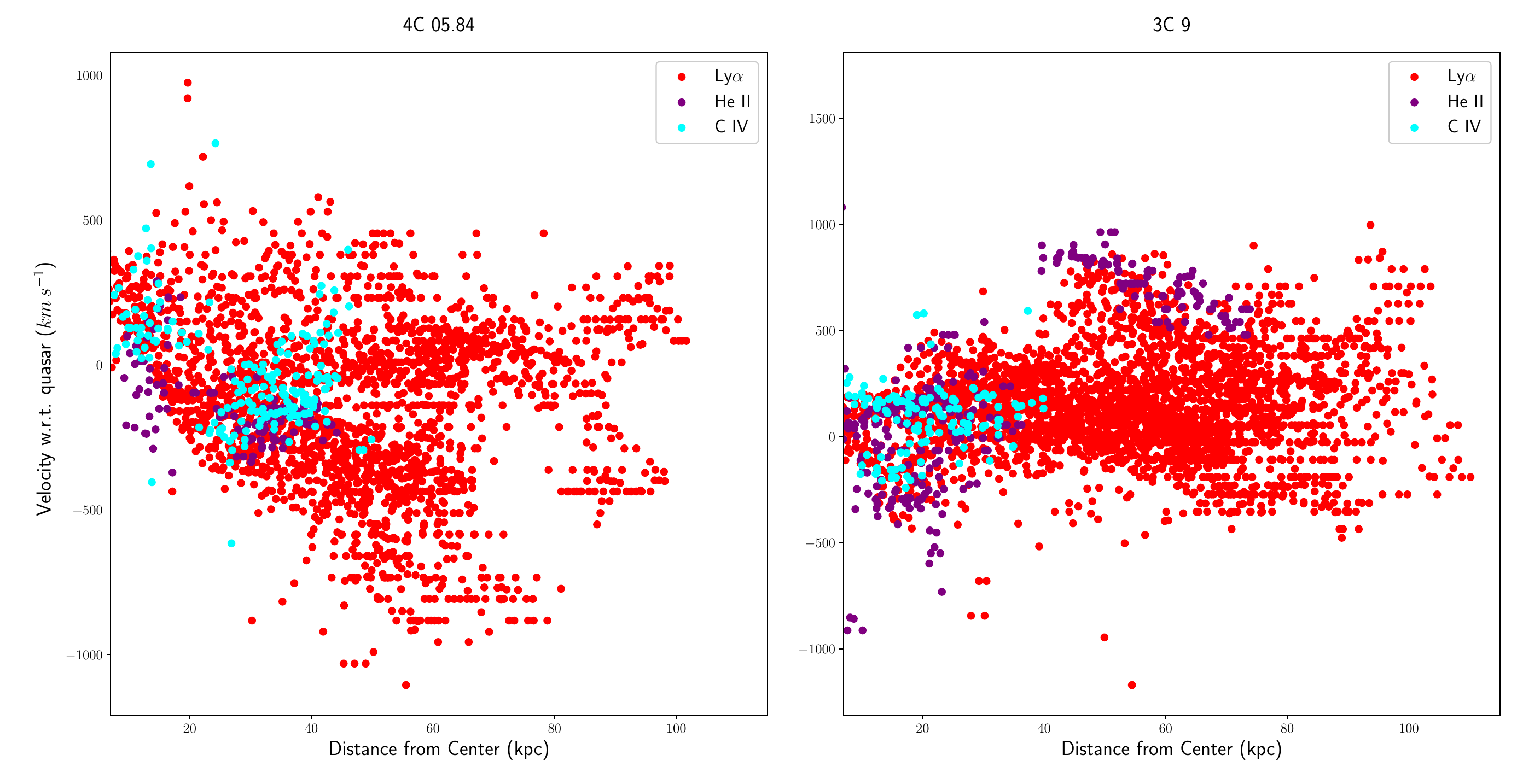} }}%
    \caption{We construct the cylindrical projection of the CGM around the two quasars by plotting the velocity with respect to the quasar against the projected distance from the quasar using data from Moment maps in Figures \ref{fig:KCWI-3C9} and \ref{fig:KCWI-4C0584}. We only present datapoints at least {7 kpc} away from the quasar since the moment maps obtained from the KCWI data may contain residuals from PSF Subtraction within this region. We can compare these plots to the cylindrical projection presented in \cite{Chen2020}}. %
    \label{fig:Cyl_Proj}
\end{figure*}

Recent observational models have shown distinct dynamical structures to individual CGMs that may be interesting to further explore observationally with KCWI. \cite{Chen2020} observed $\sim 200,000$ sightlines and absorption profiles through the intervening medium to study the CGM in aggregate and found a simple two component model worked best for their observations. This two component model has an inner and outer component. The inner CGM extends up to the inner 50 kpc and consists of outflows with velocities up to 600 \kms, whereas the outer CGM extends to 200 kpc with significantly lower velocities. Theoretically, the two component model implies that the inner CGM is the part of the CGM that is closer to the galaxy and directly encounters any galactic scale outflows, whereas the outer CGM is a larger, more diffuse body of gas. \\

To study the signatures of a potential two component model with our data, we create a scatter plot of velocity vs. distance from quasar as shown in Figure \ref{fig:Cyl_Proj} with the three emission lines we detect using KCWI, i.e. \lya, \heii\,and \civ. To generate Figure \ref{fig:Cyl_Proj} we use data in our moment 1 maps from Figures \ref{fig:KCWI-3C9} and \ref{fig:KCWI-4C0584} with the calculated the {on-sky projected} distance of each spaxel from the quasar. We then exclude any spaxels that are within the inner {0.9 arcseconds} for the quasars since these spaxels are susceptible to residuals from PSF subtraction. At the redshift of the quasars, this distance translates to about {7 kpc}. We therefore are able to probe distances larger than {7 kpc} and have the ability to resolve a potential inner CGM. \cite{Chen2020} created a similar plot but instead used the probed column densities to create a cylindrical projection of the CGM in the aggregate by combining all of their quasar sightlines. While they are able to probe the apparent optical depth of the hydrogen clouds, we do not have that information. \\

We find the \lya\, emission velocity is centered around the quasar redshift for 3C 9, whereas for 4C 05.84 the velocity is offset from the quasar center. Yet, given the resonant nature of \lya, it is difficult to deduce any evidence of coherent outflows close to the quasar, so we are only able to observe the "Outer" CGM using \lya. However, \heii\, and \civ\, provide further information about the kinematic structure of the CGM, including signatures of outflows in the inner CGM. \\

For 3C 9, we detect traces of outflows within the inner 50 kpc with velocities exceeding 500 \kms\, and approaching approximately 900 \kms. The \civ\, and \heii\, emission coincident within the inner 40 kpc are associated with the CGM of the quasar host galaxy. For 3C 9, the \heii\, emission between 40 and 80 kpc from the quasar is associated with the fringes of the detected \lya\, "Outer" CGM velocity structure as seen in Figure \ref{fig:Cyl_Proj}. We know from our MOSFIRE observations that the \heii\, in region \textit{3C9-c} is associated with companion galaxy 3C9-MOS-2. The fact that the \heii\, kinematics are on the fringes of the outer CGM highlights the stark velocity difference between the CGM around the quasar host galaxy and 3C9-MOS-2. We note that this may be the signature of the CGM from the infalling galaxy (3C9-MOS-2) starting to align with the larger CGM of the quasar host galaxy. \\

For 4C 05.84, we detect \heii\, and \civ\, coexisting with two distinct regions within the inner 40 kpc. The inner 20 kpc of 4C 05.84 \civ\, and \heii\, correspond to the quasar host galaxy, whereas the secondary structure from 20 to 40 kpc corresponds to the companion galaxies MOS-1, MOS-2 and MOS-3. We note that the observed \lya\, emission profile for 4C 05.84 in Figure \ref{fig:Cyl_Proj} has some peculiar features. Firstly, the observed \lya\, emission close to the quasar actually does not converge upon the redshift of the quasar but at a redshifted velocity of $\sim 600$ \kms\, for $z=2.32$. While we have used the value of $z=2.32$ throughout this paper from \cite{QUART2}, 4C 05.84 has earlier been reported to have $z=2.323$ \citep{Barthel1990}. If taken to be the redshift of the quasar, the observed CGM emission would change the velocity offset to $\sim 300$ \kms. The other noteworthy observation is the large difference in velocities between the "inner" CGM and the \lya\, emission 40-80 kpc away where we see a systematic blueshift in \lya\, emission, offset by as much as 1000 \kms\, compared to the inner CGM velocity. \\

sWe now compare the observed two component structure for the two quasars 3C 9 and 4C 05.84. We notice that while for 3C 9, the larger \lya\, emission neatly converges towards the redshift of the quasar, for 4C 05.84, the emission close to the quasar seems to be offset from the mean velocity of \lya\, emission for distances larger than 20 kpc. This indicates the host galaxy may have experienced a shift in velocity compared to the larger CGM around it in the past. This is interesting since \cite{QUART2} did not detect any signatures of a merger in diffraction limited observations of the host galaxy ISM kinematics. Additionally, the large velocity offset between the innermost and outer parts of the CGM in 4C 05.84 is reminiscent of the \heii\, emission joining the outer CGM in 3C 9 between 40-80 kpc as seen in Figure \ref{fig:Cyl_Proj}. If the CGM is indeed a two component model where the outer halo is largely confined to $\pm 500$ \kms\, for distances larger than 10-20 kpc, then this \lya\, emission could be indicative of gas accreting on to the outer CGM from the cosmic web. \\

In summary, we detect signatures of outflow consistent with those expected from the inner and outer CGM for both 3C 9 and 4C 05.84. We detect potential outflows from the inner CGM using \heii\, emission, whereas the outer CGM is detected and studied using \lya, \heii\, and \civ.

\subsection{{Source of CGM ionization}}\label{subsec:CGM_ionization_metallicity}

\begin{figure*}[!ht]
	\centering
	{{\includegraphics[width=\linewidth]{ 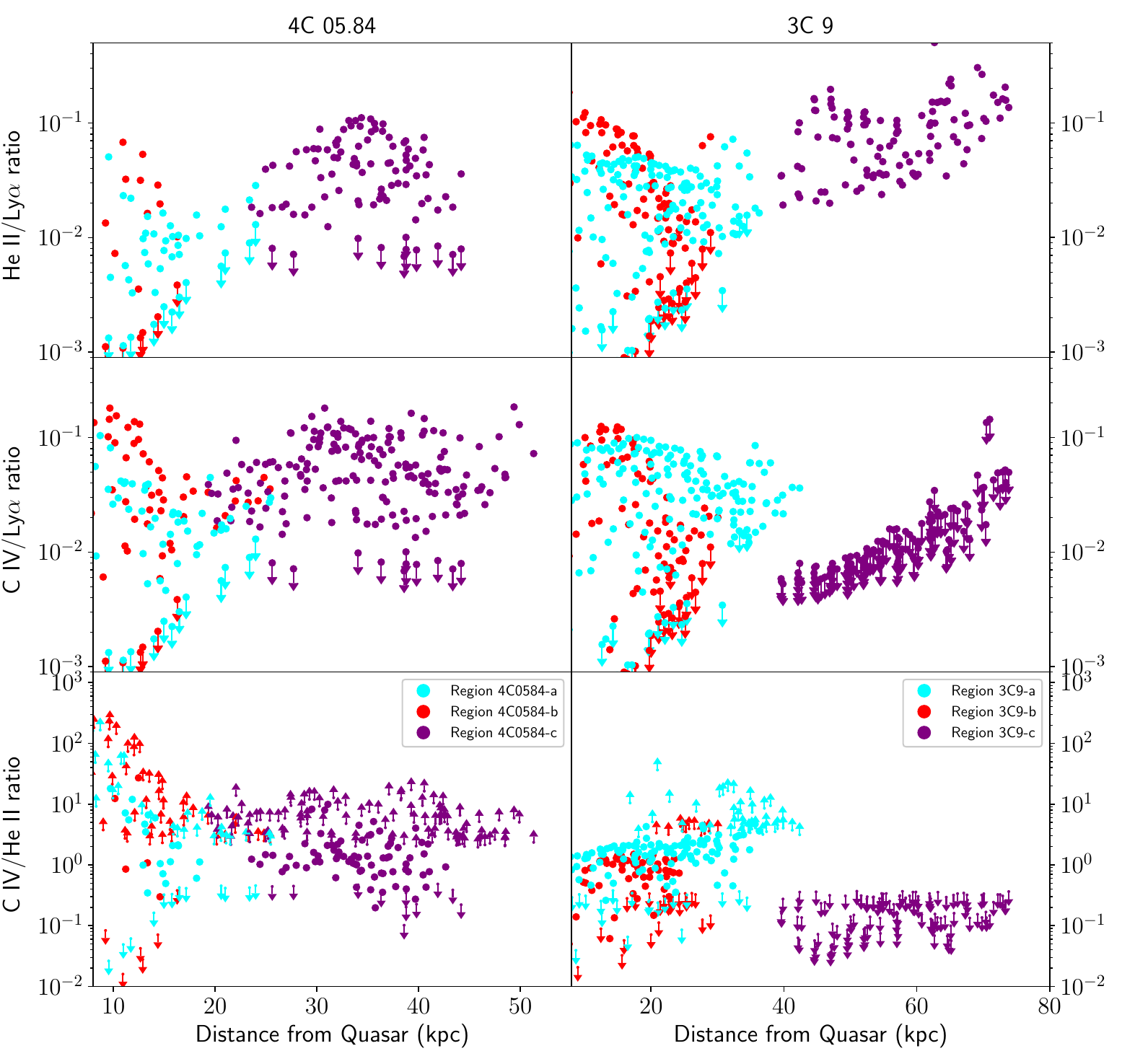} }}%
	\caption{We investigate sources of ionization in the CGM by looking at \heiilya, \civlya\, and \civ/\heii\, ratios as a function of distance from the quasar. The 4 inner regions \textit{3C9-a}, \textit{3C9-b}, \textit{4C0584-a} and \textit{4C0584-b} all have radially decreasing \heiilya\, and \civlya\, ratios, consistent with quasar photoionization. Regions \textit{3C9-c} and \textit{4C0584-c} do not show such a radial trend, with the \heiilya\, ratio for \textit{4C0584-c} peaking at a distance of 32 kpc from the quasar, whereas for \textit{3C9-c}, it decreases and then increases. Regions \textit{4C0584-a} and \textit{4C0584-b} also show a radially decreasing \civ/\heii\, ratio, whereas 3C 9 does not show such a trend. These trends are consistent with the picture that Regions \textit{3C9-a}, \textit{3C9-b}, \textit{4C0584-a} and \textit{4C0584-b} are part of the quasar host galaxy CGM and ionized by the quasar, whereas Regions \textit{3C9-c} and \textit{4C0584-c} are associated with the CGM surrounding companion galaxies.} %
	\label{fig:Ratios_Radial}
\end{figure*}

{We use the observed \lya, \heii\, and \civ emission profiles to discuss the source of ionization powering the CGM emission. Figure \ref{fig:Ratios_Radial} shows the \heiilya, \civlya\, and \civ/\heii\, ratios as a function of radial on-sky separation from the quasar. We have only plotted spaxels at a distance greater than 8 kpc to remove any spaxels with PSF subtraction residuals. While \heii\, and \civ\, have been used to distinguish AGN ionization from shocks in the ISM, they require the \ciii\, line \citep{VillarMartin1997, Nagao2016, Feltre2016, Humphrey2019}. Instead, we use the observed \heiilya\, and \civlya\, ratios to consider the source of ionization for the observed CGM profiles around these two quasars. \\
	Firstly, we note that the observed \heiilya\, ratios for both quasars are less than 0.15 for all observed spaxels with significant detection. \cite{Cantalupo2019} obtained theoretical limits on \heiilya\, ratios of 0.23 and 0.30 for Case A and Case B recombination respectively at T$ = 2 \times 10^4$ K, for fully ionized \heii, suggesting that the \heii\, is not fully ionized.\\ 
	Secondly, we note that for Regions \textit{3C9-a}, \textit{3C9-b}, \textit{4C0584-a} and \textit{4C0584-b}, corresponding to CGM gas directly associated with the quasar host galaxy, the observed \heiilya\, and \civlya\, ratios, wherever data is available, decrease with greater on-sky separation from the quasar. Comparing our results to photoionization models in \cite{Humphrey2019}, we find that the decreasing ratios are consistent with a decrease in \logU. A radially decreasing \logU\, parameter is consistent with quasar photoionization. Regions \textit{3C9-c} and \textit{4C0584-c}, which are likely associated with companion galaxies, do not show radially declining ratios. Region \textit{4C0584-c} shows a peak at $\sim$ 32 kpc from the quasar for the \heiilya\, ratio, whereas \textit{3C9-c} shows a radial decrease, followed by an increase. This could be explained by line of sight projection effects. If we assume that the CGM around the companion galaxies is ionized by the quasars and the CGM and companion galaxies are extended along the line of sight, then the radial distance from the quasar will change significantly, compared to the on-sky projected separation. For these regions, we require the \ciii\, line to definitively identify the source of ionization.\\
	Thirdly, we compare our results to those of \cite{Guo2020} obtained from stacking $z=3$ CGM spectra from 80 quasars observed with MUSE. Since we do not have data for the \ciii\, line, we consider two regimes, whether \civ/\heii\, is greater than or less than 1. 
	We consider the case where \civ/\heii\, ratios are less than 1, relevant for the region \textit{3C9-c}. Based on the photoionization modelling in \cite{Guo2020}, this would suggest \logU$<2.4$\, in this region, whereas no constraints can be placed on the metallicity based on this line ratio alone. This is again consistent with quasar photoionization. However, since \textit{4C0584-c} and \textit{3C9-c} are at comparable on-sky projected distances from the quasar, the lack of \civ\, detected in \textit{3C9-c} could indicate a significantly larger distance from the quasar along the line of sight, resulting in significantly lower \logU\, parameter value. The other explanation for a lack of \civ\, detection is if the observed gas in Region \textit{3C9-c} is an infalling subhalo around the companion galaxy, the companion galaxy has not undergone significant star formation prior to infall, and thus has not formed and ejected metals into the CGM, resulting in a low metallicity. However, confirmation of this scenario would require additional diagnostics for the CGM. \\
	Lastly, we consider the biconical ionization around both quasars as seen in \heii\, and \civ. Since these likely indicate quasar photoionization cones, we consider the implications for the larger \lya\, nebula. The presence of these biconical regions around both quasars, and the presence of the extended regions \textit{4C0584-c} and \textit{3C9-c} within the opening angles of these cones strongly indicate \civ\, and \heii\, emission in these regions being powered by quasar photoionization. The question then arises: Is this also true for the \lya\, emission? If only the inner CGM is directly photoionized by the quasar, this would suggest that recombination may not be the only significant mechanism powering \lya\, emission. Further confirmation of this scenario would require \ciii\ observations with additional radiative transfer modelling for these systems. We also note the important caveat that radio jets can affect the observed lines, since they can modify the gas densities and kinematics in these systems. 
}

\section{Summary} \label{sec:Summary}
We have presented results for two radio-loud quasars, 3C 9 and 4C 05.84. By combining optical observations of the CGM with near-infrared spectroscopy of companion sources, we have comprehensive picture of the environment around these quasars. While previous studies have investigated diffuse gas around radio-loud quasars at $z \sim 2$ using long slit spectroscopy \citep{Shukla2022}, or explored \lya\, nebulae around radio-loud quasars at $z=3$ \citep{Borisova2016,AB2019}, our work stands out as one of the initial multi-wavelength analysis of \lya\, nebulae $z=2$ radio-loud quasars with IFU observations. We have uniquely identified companion galaxies that potentially contribute to observed \lya\, nebulae emission and have conducted a comparative analysis of the nebulae's orientation in relation to the radio jets. We summarise our findings as follows:
\begin{enumerate}
    \item For both quasars, we detect large \lya\,nebulae, projected on sky to be {120 kpc} across, with spatially resolved substructure down to a sensitivity of {$5 \times 10^{-20}$\fergarc}. 
    \item We detect \heii\,and \civ\, nebulae around these quasars about {40-50 kpc} across as projected on sky. While we detect these nebulae immediately around the quasar, we also detect {$\sim 20$ kpc diameter} kinematically distinct regions 15-30 kpc away from both quasars. 
    \item The \heii\,nebulae around both quasars show a biconical shape, indicating that they are being powered due to ionization by the quasar. The shape of the nebulae traces this ionization cone which further strengthens the fact that the detected \lya\, and \heii\, nebulae immediately around the quasar are aligned with their observed radio jets. {Observed \civlya\, and \heiilya\, ratios additionally show that the CGM emission is predominantly powered by quasar photoionization. This assertion is strengthened by the biconical morphology of \heii\, and \civ\, emission seen around both quasars, indicative of an ionization cone.}
    \item We detect \ha, \hb\, and \oiii\ emission {from the companion galaxies} in the environment around each quasar using MOSFIRE observations. For 3C 9, we detected \ha\, and \oiii\, coincident with the kinematically distinct southern \heii\, region \textit{3C9-c} indicating the presence of a companion galaxy. We also detect \ha\, emission in a galaxy about 172 kpc away from the quasar.
    \item {The observed radial profile for \lya\, emission around these two quasar host galaxies deviates from a smooth gradual decline. These surface brightness deviations align with the locations of \lya, \heii\, and \civ\, emission from the CGM likely associated with the companion galaxies. This association is confirmed by both the spatial and velocity match of the observed \ha, \oiii\, and \hb\, emission. Therefore, the observed substructure in \lya, nebular emissions around high-redshift quasars may serve as an indicator of the presence of companion galaxies.}
    \item {Confirmed companion galaxies around 3C 9 shows this source is in active cluster formation. The mass of the 3C 9 cluster progenitor is estimated to be $2.7 \times 10^{12}$ \msun. Confirmed companion sources around the 4C 05.84 were similarly used to calculate the mass of the group progenitor to be $3.5 \times 10^{12}$ \msun. We estimate that the total mass of the 4C 05.84 system will grow to about $2.3 \times 10^{13} \, \mathrm{M}_{\odot}$ by present day whereas 3C 9 will grow to a mass of $2.15 \times 10^{14} \, \mathrm{M}_{\odot}$ by present day. Assuming that these structures are not yet gravitationally bound, this makes the detected structure around 4C 05.84 a protogroup and that around 3C 9 a protocluster. }
\end{enumerate}
\section*{Acknowledgements}
 The authors wish to thank Sherry Yeh, Josh Walawender, Carlos Alvarez, Percy Gomez (W. M. Keck Observatory support astronomers) and Julie, Heather, John, Tony C. and Luca (Observing Assistants) with their assistance at the telescope to acquire the Keck MOSFIRE and KCWI data sets. The authors thank Professor Karin Sandstrom for valuable discussions surrounding nebular line ratio calculations. The authors also wish to thank the anonymous referee for their valuable inputs and suggestions. The data presented herein were obtained at the W. M. Keck Observatory, which is operated as a scientific partnership among the California Institute of Technology, the University of California and the National Aeronautics and Space Administration. The Observatory was made possible by the generous financial support of the W.M. Keck Foundation. The authors wish to recognize and acknowledge the very significant cultural role and reverence that the summit of Maunakea has always had within the indigenous Hawaiian community. We are most fortunate to have the opportunity to conduct observations from this mountain. \\
 This publication makes use of data products from the Two Micron All Sky Survey, which is a joint project of the University of Massachusetts and the Infrared Processing and Analysis Center/California Institute of Technology, funded by the National Aeronautics and Space Administration and the National Science Foundation. This research has made use of the Keck Observatory Archive (KOA), which is operated by the W. M. Keck Observatory and the NASA Exoplanet Science Institute (NExScI), under contract with the National Aeronautics and Space Administration. Some of the data presented in this paper were obtained from the Mikulski Archive for Space Telescopes (MAST). STScI is operated by the Association of Universities for Research in Astronomy, Inc., under NASA contract NAS5-26555. Support for MAST for non-HST data is provided by the NASA Office of Space Science via grant NNX13AC07G and by other grants and contracts. This work is based [in part] on observations made with the Spitzer Space Telescope, which is operated by the Jet Propulsion Laboratory, California Institute of Technology under a contract with NASA.\\
 This research made use of Astropy, a community-developed core Python package for Astronomy \citep{2013A&A...558A..33A} 
\bibliography{3c9_KCWI_MOSFIRE}{}
\bibliographystyle{aasjournal}

\appendix
\section{MOSFIRE and KCWI 1D Spectra}
1D spectra extracted from KCWI and MOSFIRE for all identified sources around the two quasars are presented here. The units and methodology for figure preparation is identical to those used for Figures \ref{fig:MOS-4C0584} and \ref{fig:MOS-3C9}.
\subsection{3C 9}

\begin{figure}[H]
	\includegraphics[width=\linewidth]{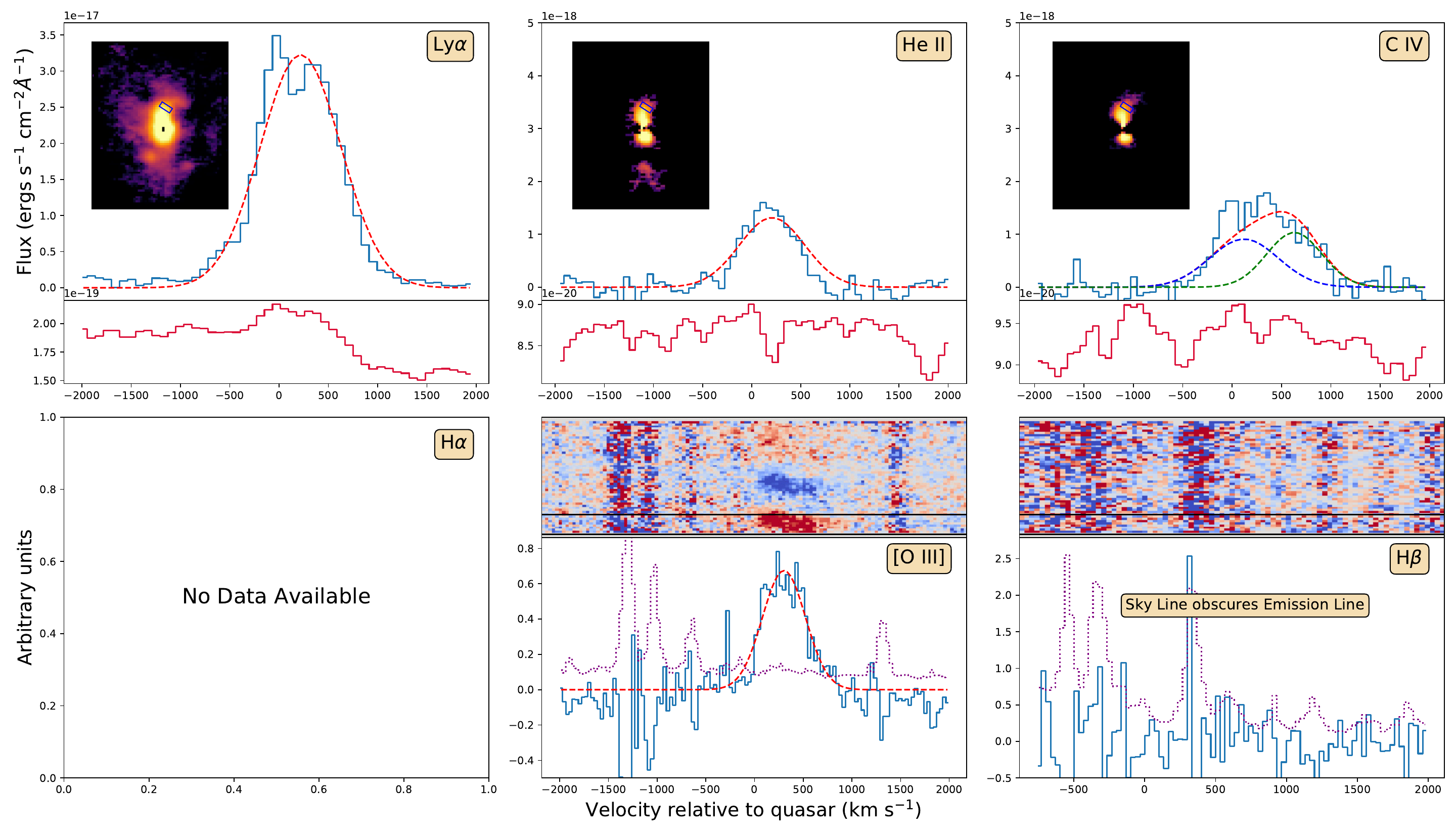}
	\caption{All spectra for 3C9-MOS-1}
	\label{fig:3C9-MOS-1}
\end{figure}

\begin{figure}[H]
	\includegraphics[width=\linewidth]{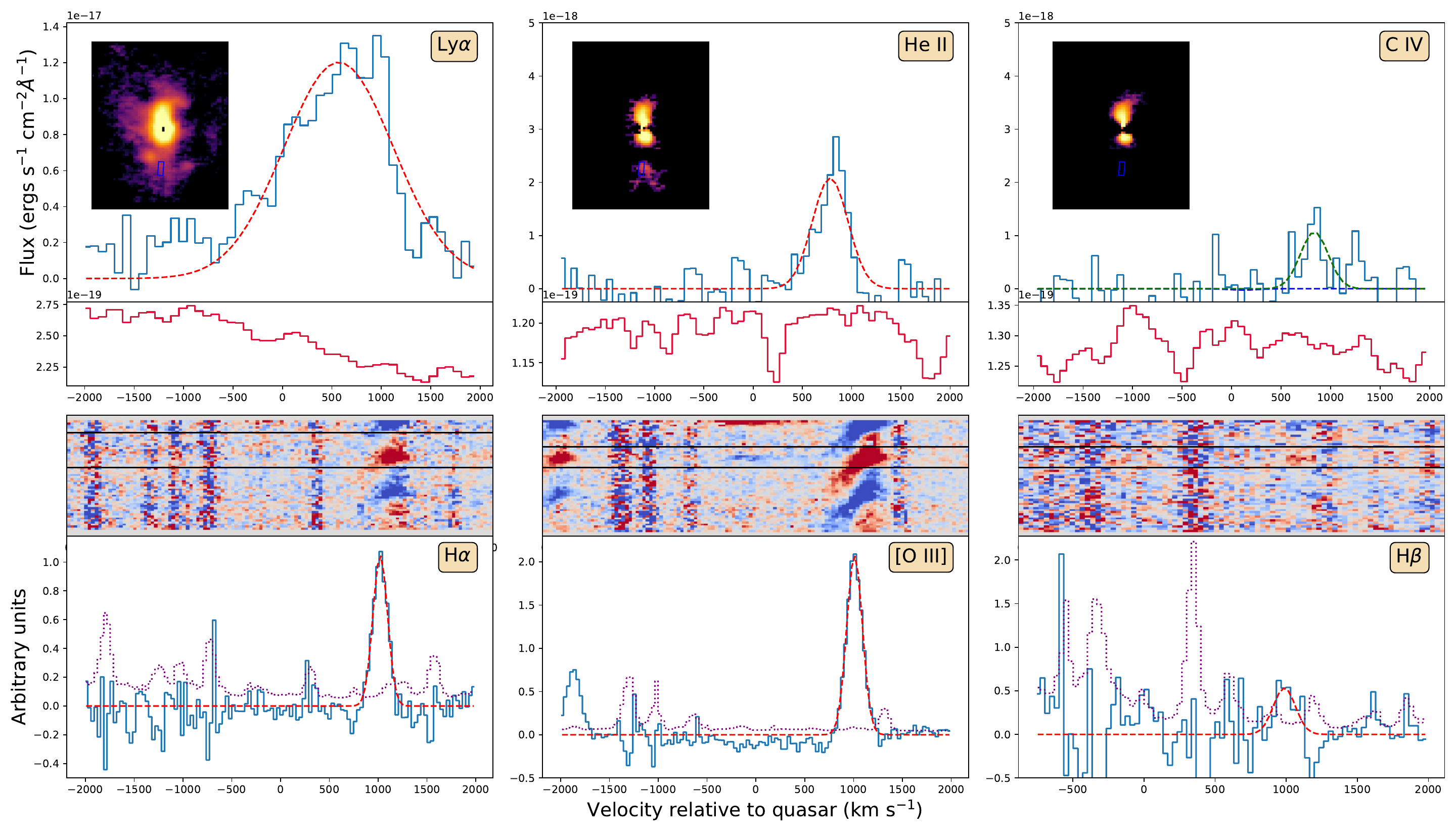}
	\caption{All spectra for 3C9-MOS-2}
	\label{fig:3C9-MOS-2}
\end{figure}

\begin{figure}[H]
	\includegraphics[width=\linewidth]{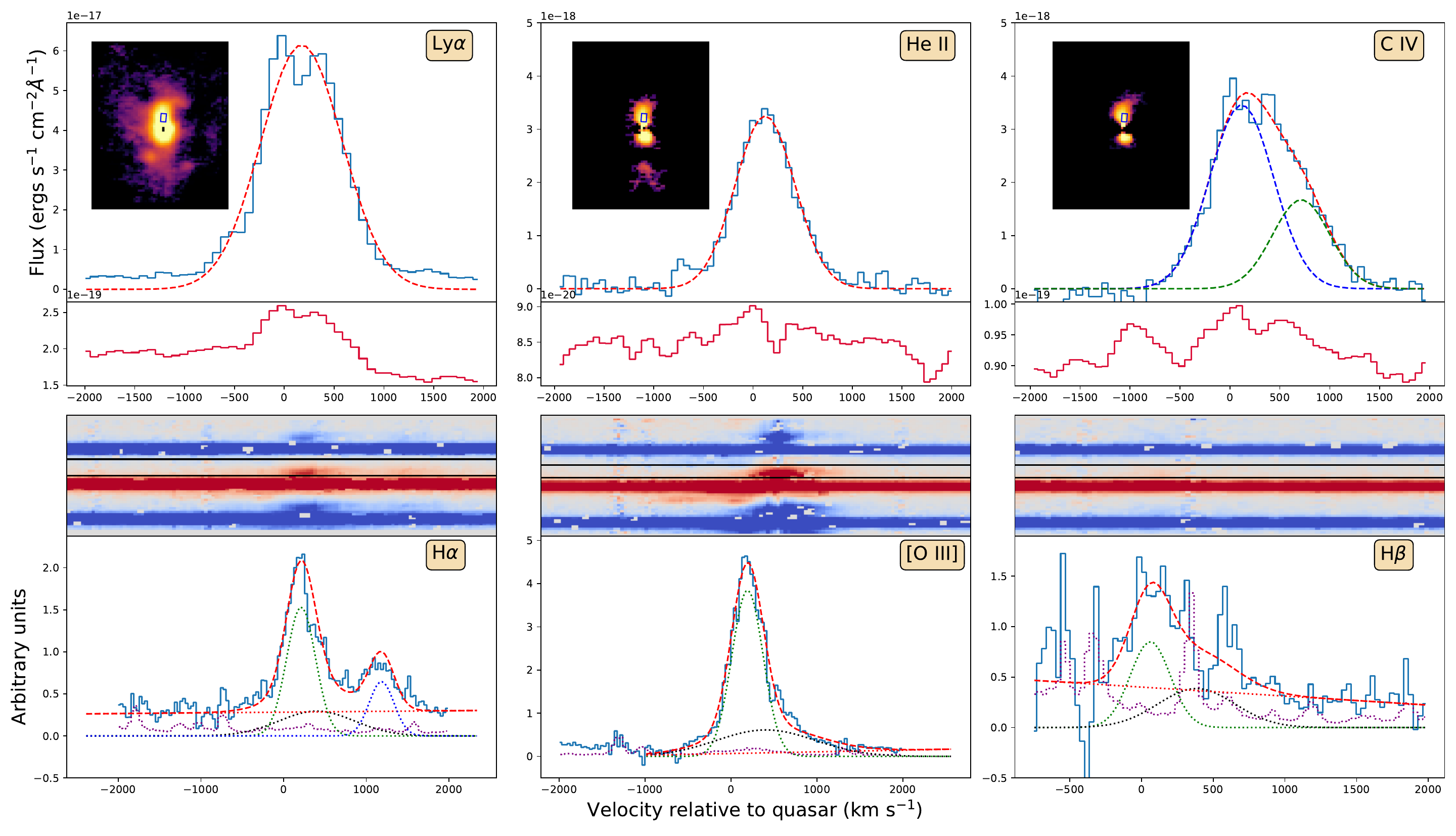}
	\caption{All spectra for 3C9-MOS-3}
	\label{fig:3C9-MOS-3}
\end{figure}

\begin{figure}[H]
	\includegraphics[width=\linewidth]{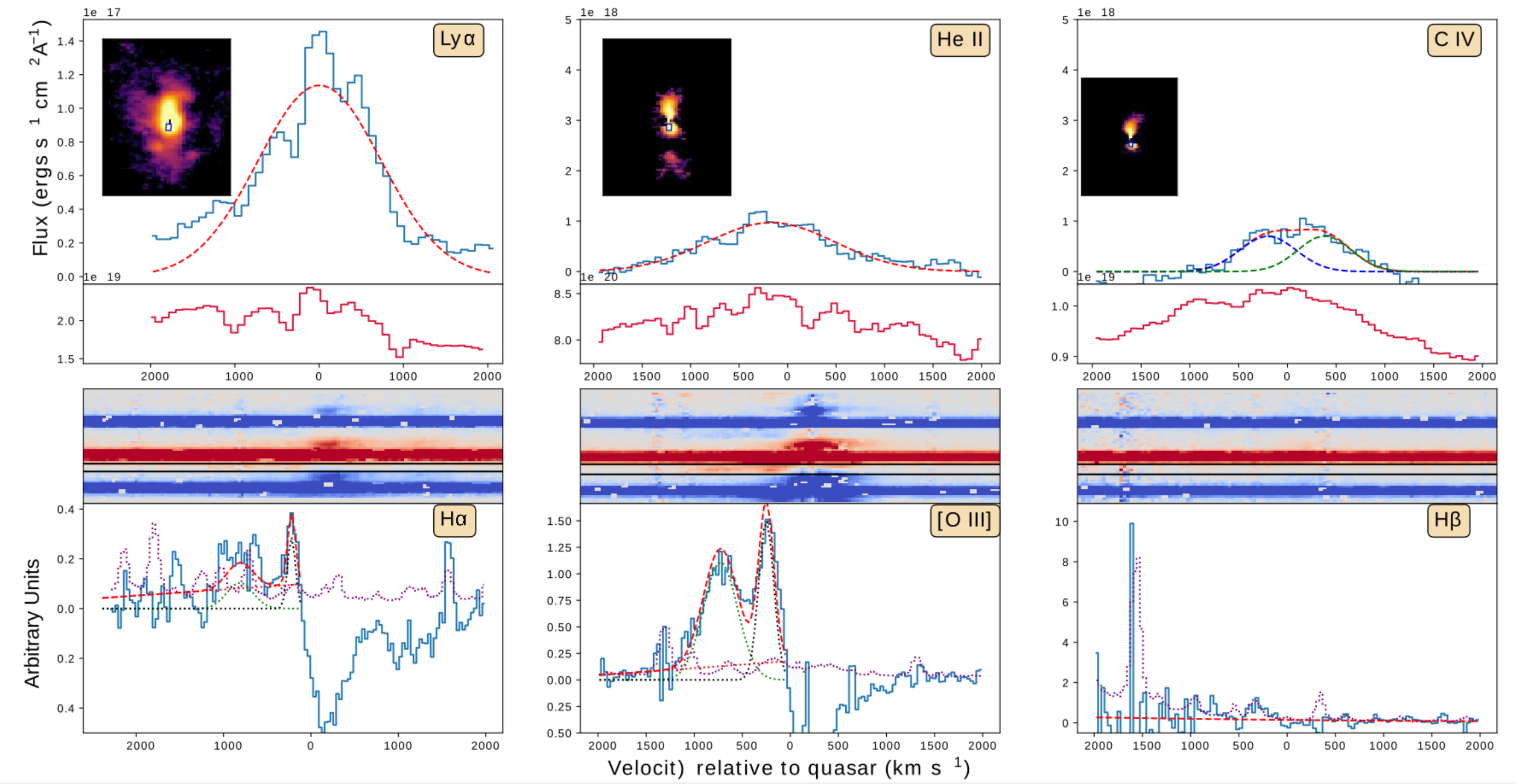}
	\caption{All spectra for 3C9-MOS-4}
	\label{fig:3C9-MOS-4}
\end{figure}

\begin{figure}[H]
	\includegraphics[width=\linewidth]{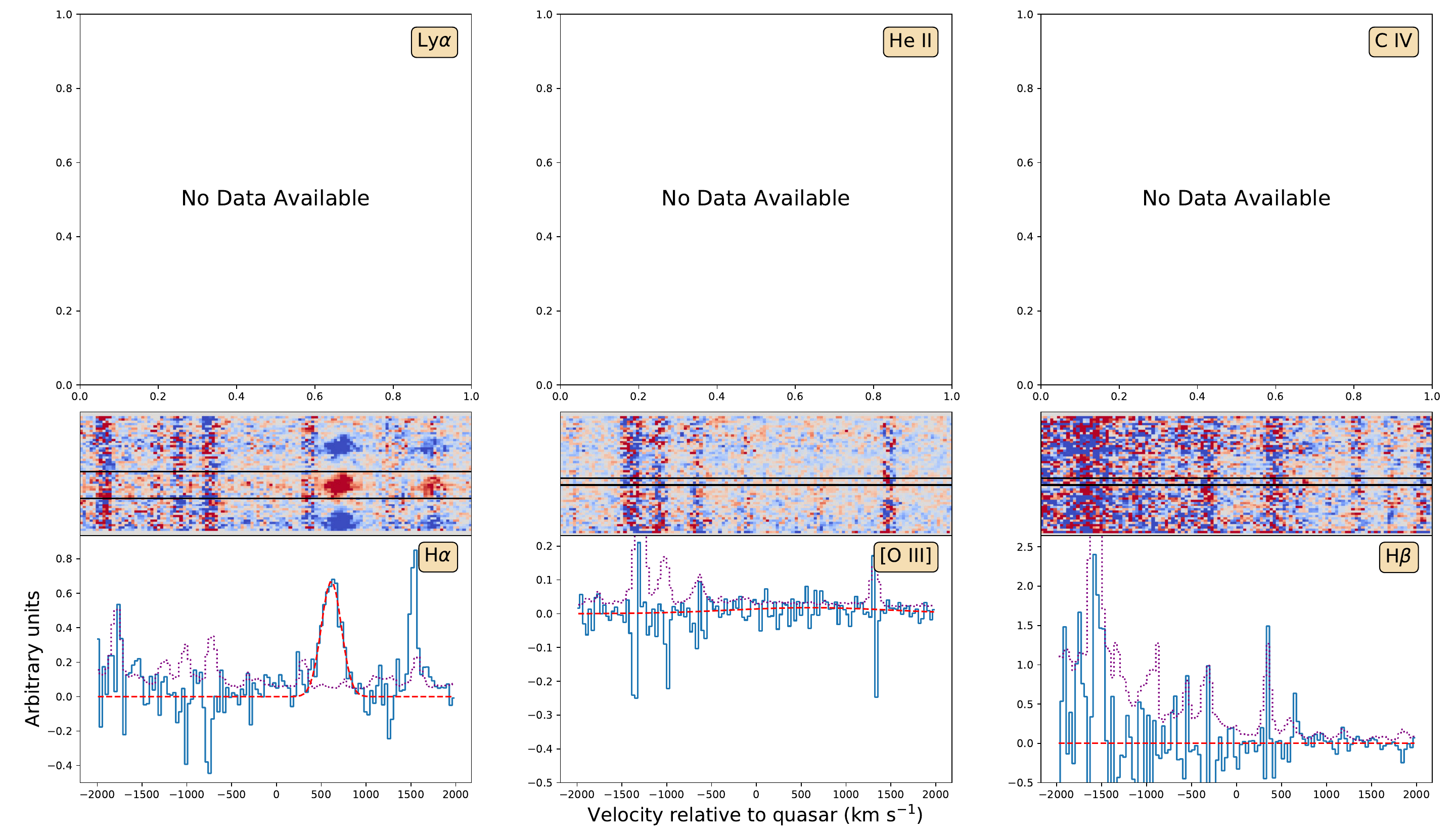}
	\caption{All spectra for 3C9-MOS-5}
	\label{fig:3C9-MOS-5}
\end{figure}

\pagebreak

\subsection{4C 05.84}

\begin{figure}[ht!]
	\includegraphics[width=\linewidth]{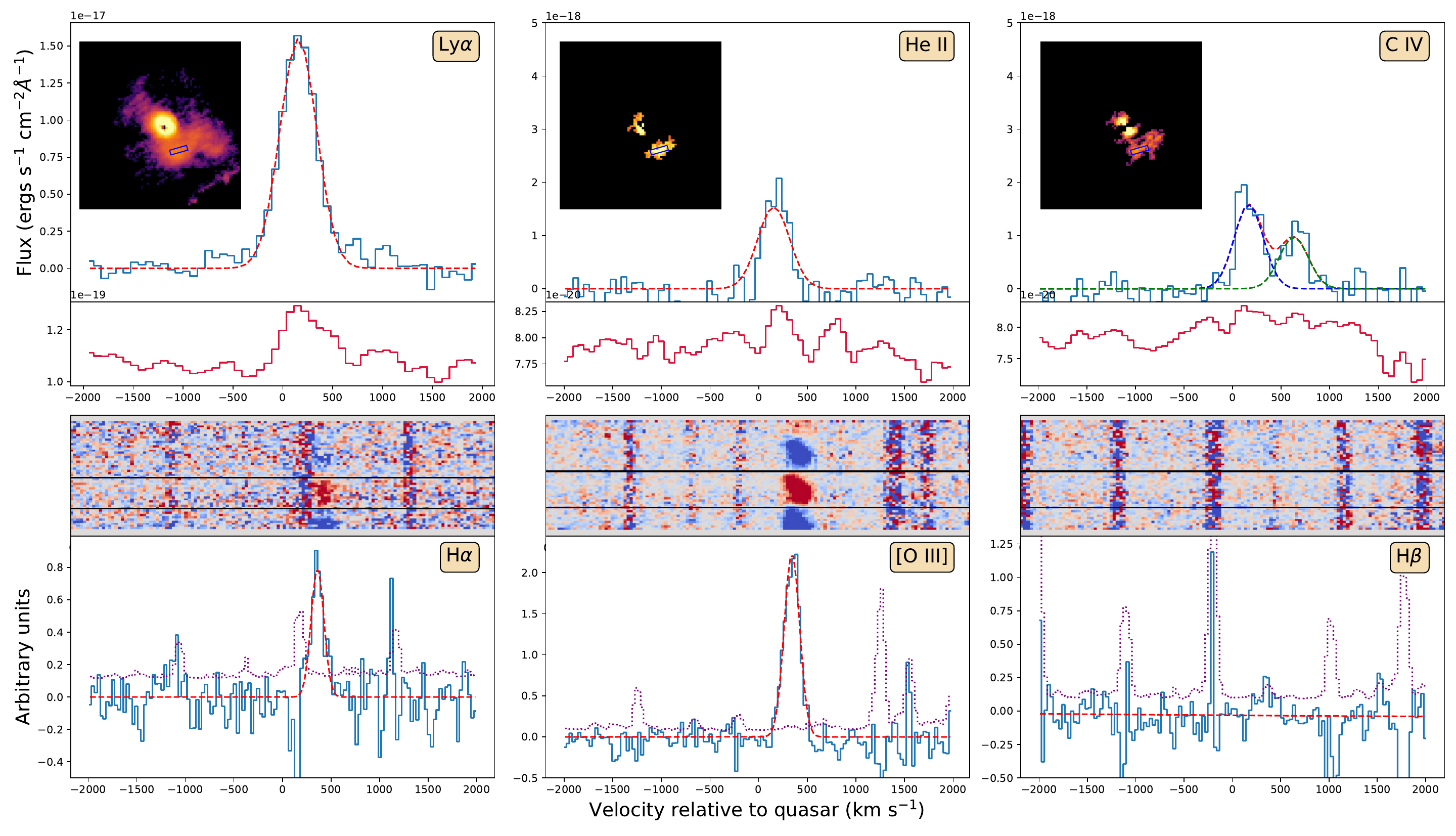}
	\caption{All spectra for 4C0584-MOS-1}
	\label{fig:4C0584-MOS-1}
\end{figure}

\begin{figure}[H]
	\includegraphics[width=\linewidth]{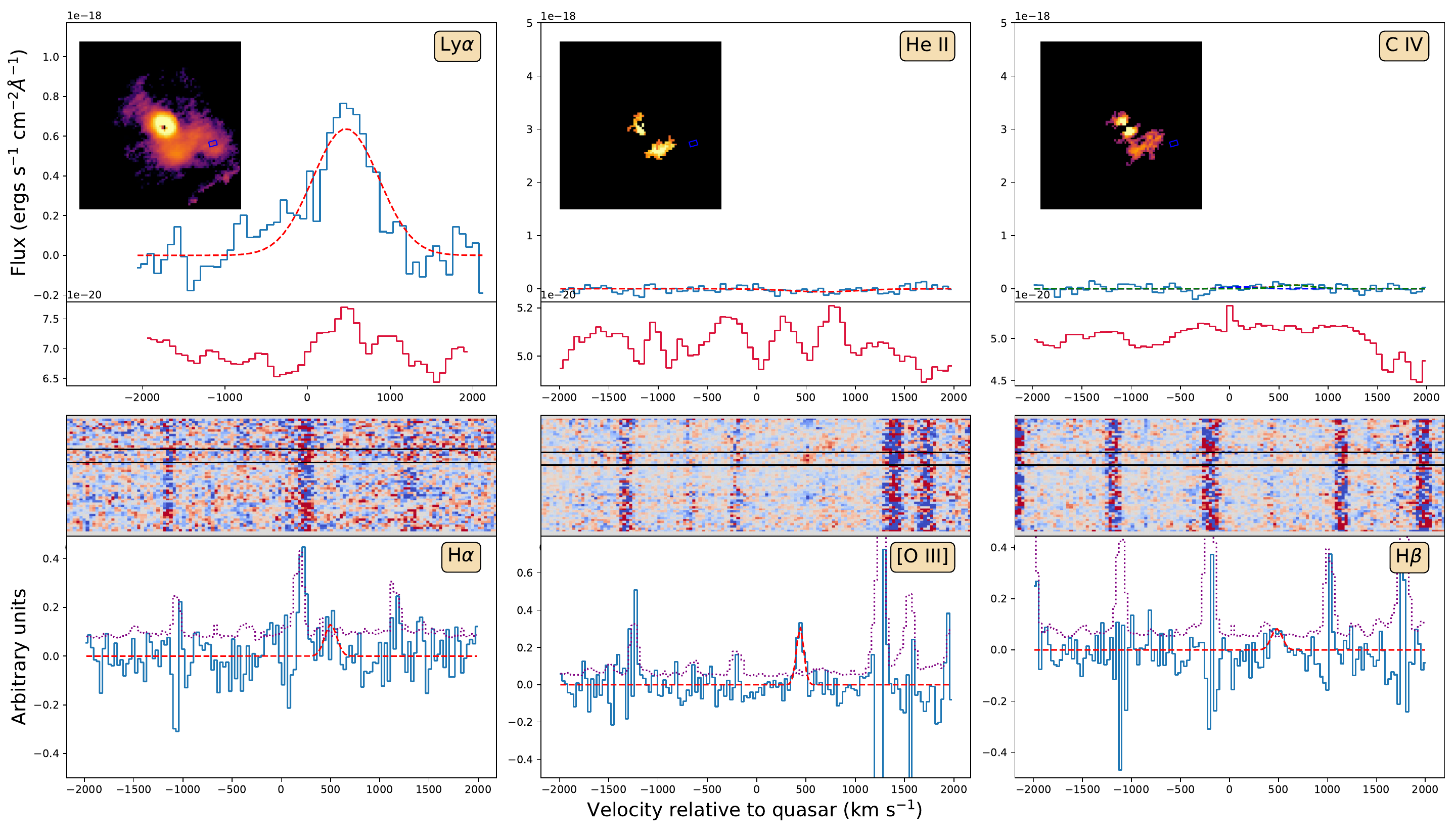}
	\caption{All spectra for 4C0584-MOS-2}
	\label{fig:4C0584-MOS-2}
\end{figure}

\begin{figure}[H]
	\includegraphics[width=\linewidth]{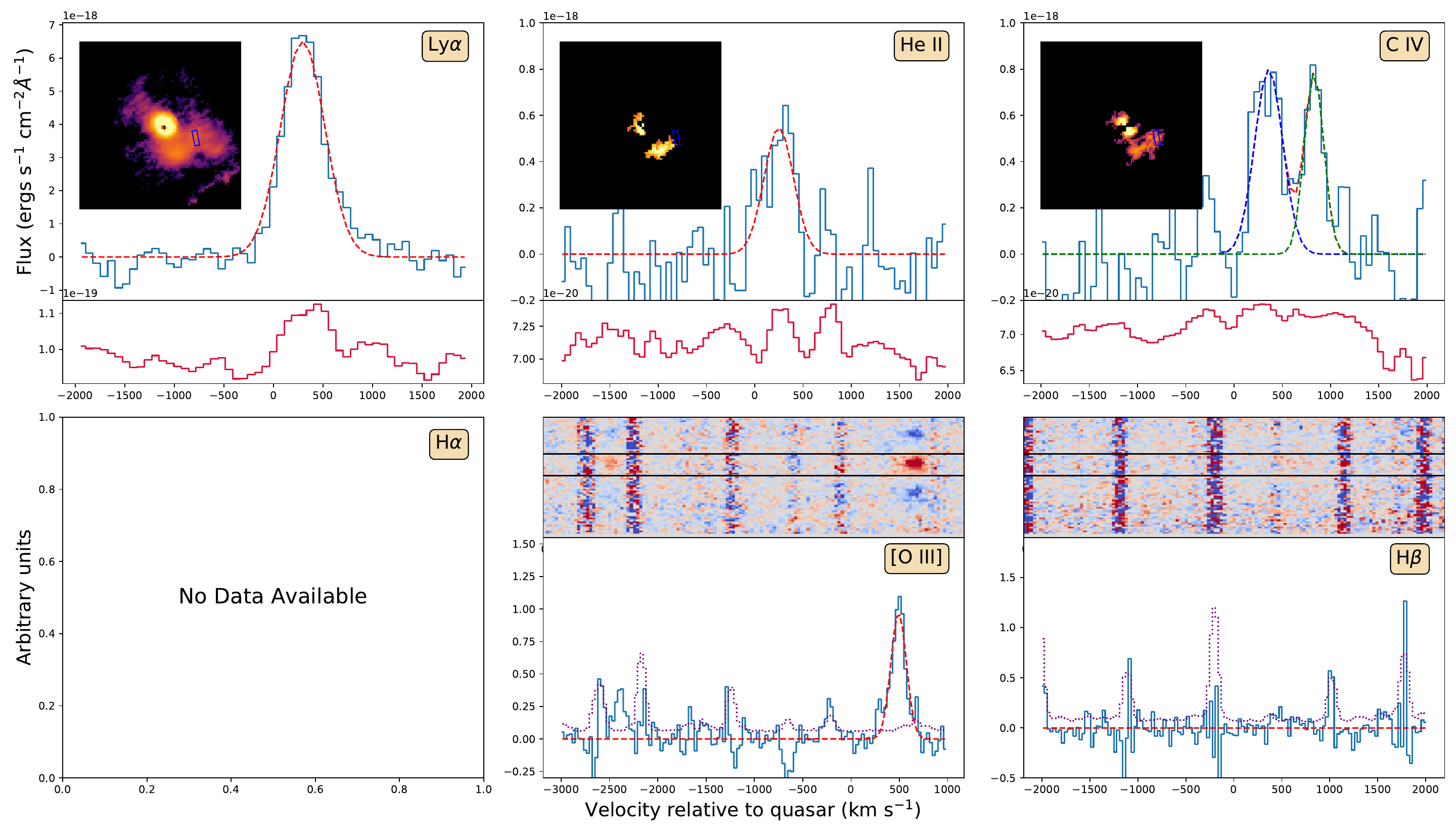}
	\caption{All spectra for 4C0584-MOS-3}
	\label{fig:4C0584-MOS-3}
\end{figure}

\end{document}